\documentclass{emulateapj}
\usepackage{color}
\definecolor{note_fontcolor}{rgb}{0.80078125, 0.80078125, 0.80078125}
\usepackage{graphicx}

\slugcomment{}
\shorttitle{}
\shortauthors{}

\begin{document}

\title{\emph{The Carnegie Supernova Project}: Intrinsic Colors of Type Ia Supernovae}

\author{
Christopher R. Burns\altaffilmark{1},
Maximilian Stritzinger\altaffilmark{2},
M. M. Phillips\altaffilmark{3},
E. Y. Hsiao\altaffilmark{3},
Carlos Contreras\altaffilmark{3,2},
S. E. Persson\altaffilmark{1},
Gaston Folatelli\altaffilmark{4},
Luis Boldt\altaffilmark{3},
Abdo Campillay\altaffilmark{3},
Sergio Castell\'on\altaffilmark{3},
Wendy L.\ Freedman\altaffilmark{1},
Barry F. Madore\altaffilmark{1},
Nidia Morrell\altaffilmark{3},
Francisco Salgado\altaffilmark{3} and
Nicholas B. Suntzeff\altaffilmark{5}
}
\altaffiltext{1}{Observatories of the Carnegie Institution for Science, 813 Santa Barbara St, Pasadena, CA, 91101, USA}
\altaffiltext{2}{Department of Physics and Astronomy, Aarhus University, Ny Munkegade 120, DK-8000 Aarhus C, Denmark}
\altaffiltext{3}{Carnegie Institution of Washington, Las Campanas Observatory, Colina El Pino, Casilla 601, Chile}
\altaffiltext{4}{Kavli Institute for the Physics and Mathematics of the Universe, Todai Institutes for Advanced Study, the University of Tokyo, 277-8583 Kashiwa, Japan}
\altaffiltext{5}{George P. and Cynthia Woods Mitchell Institute for Fundamental Physics and Astronomy, Texas A\&M University, Department of Physics and Astronomy, College Station, TX, 77843, USA}

\begin{abstract}
We present an updated analysis of the intrinsic colors of SNe~Ia using
the latest data release of the \emph{Carnegie Supernova Project}. We introduce
a new light-curve parameter very similar to stretch that is better
suited for fast-declining events, and find that these peculiar types
can be seen as extensions to the population of ``normal'' SNe~Ia.
With a larger number of objects, an updated fit to the
Lira relation is presented along with evidence for a dependence on the late-time
slope of the $B-V$ light-curves with stretch and color. Using the
full wavelength range from $u$ to $H$ band, we place constraints
on the reddening law for the sample as a whole and also for individual
events/hosts based solely on the observed colors. The photometric
data continue to favor
low values of $R_{V}$, though with large variations from event
to event, indicating an intrinsic distribution. We confirm the findings
of other groups that there appears to be a correlation between the
derived reddening law, $R_{V}$, and the color excess, $E(B-V)$, such
that larger $E(B-V)$ tends to favor lower $R_{V}$. The 
intrinsic $u$-band colors
show a relatively
large scatter that cannot be explained by
variations in $R_{V}$ or by the \citet{Goobar:2008} power-law for
circumstellar dust, but rather is correlated with spectroscopic
features of the supernova and is therefore likely due to
metallicity effects.
\end{abstract}

\keywords{distance scale --- dust, extinction --- galaxies: ISM --- methods: statistical
--- supernovae: general}

\section{Introduction}

Type Ia Supernova (SNe~Ia) cosmology is embarking on the next generation of experiments.
With the advent of near-term dark-energy missions such as the Dark Energy Survey
(DES), and longer-term projects like Euclid
and WFIRST, the number of high-redshift SNe~Ia
will exceed the low-redshift sample by almost two orders of magnitude.
These experiments are firmly in the regime where random errors such
as photometric precision, unknown SN~Ia to SN~Ia variations, and even
larger errors due to photometric redshifts and typing, will contribute
less to the error budget of the cosmological parameters than the systematic
errors. 

Among the most vexing of these systematics is the source of the observed
color distribution of SNe~Ia. As with all standard candles, SNe~Ia
are known to suffer from extinction along the line of
sight. Early work done to standardize SNe~Ia assumed that inter-stellar
dust in the Milky Way and host galaxy were primarily responsible for
making some objects redder than others \citep[e.g.][]{Phillips:1993}.
And while such treatment led
to great successes including the discovery of Dark Energy, it was not
long before inconsistencies arose, the most immediately obvious being
the abnormally low value of the ratio of total-to-selective absorption,
$R_{V}$ \citep{Tripp:1998}. Further muddying the issue is the possibility that different
sub-types of SNe~Ia may have different intrinsic colors \citep{Foley:2011b},
that the low $R_{V}$ is due to the presence of a circumstellar medium
(CSM) around the progenitor system \citep{Wang2005,Goobar:2008},
or the angle from which we view the explosion \citep{Foley:2011b,Maeda:2011}.
Most recently, work by \citet{Phillips:2013} shows that while a large
number of SNe~Ia do not follow the correlation between extinction and
\ion{Na}{1} column density seen in the Milky Way, diffuse interstellar bands
do show such a correlation, leading to the conclusion that most of
the source of reddening is interstellar in origin.

On the one hand, we can take an agnostic approach and simply find
color corrections that produce the best possible distances by
minimizing residuals in the Hubble diagram, for instance. This is unsatisfying,
however, since we cannot know to what extent, if any, these color
corrections evolve with redshift, leaving us with a systematic error
that is hard to quantify. Furthermore, most SN~Ia cosmological analyses
\citep[e.g.][]{Hicken:2009,Freedman:2009,Conley:2011} assume that there is
only one universal value for $R_{V}$
(or equivalently, one universal luminosity-color correction factor  $\beta$
\citep{Tripp:1998,Astier:2006}),
and therefore observing an increasing
number of SNe~Ia will tend to reduce the systematic error in $R_{V}$.
In reality, there is an observed intrinsic distribution of $R_{V}$
in our Milky Way (MW), representing a random error
in cosmological analyses which is not be reduced by simply increasing
the sample size of SNe~Ia.

Alternatively, one can assume that the colors of SNe~Ia arise from
both an intrinsic mechanism that correlates with physical
properties of the SN~Ia and also from interstellar dust, both
in the host galaxy and in our own Milky Way. One could argue that
in order to constrain any possible evolution of SNe~Ia, one needs
to understand, or at least quantify, the intrinsic and extrinsic reddening
mechanisms. With these issues in mind, the {\em Carnegie Supernova Project}
\citep[CSP][]{Hamuy:2006}
has observed SNe~Ia in a wide range of filters from the near ultra-violet to
the near infra-red (NIR). This large baseline in wavelength allows us to
study $R_{V}$ from SN~Ia to SN~Ia \emph{based
on colors alone}.

Another problem, at least in the sense of using SNe~Ia as standard
candles, is the large variety within the Type Ia class. Early on, it was
evident that there were at least three sub-classes: the ``normal''
SNe~Ia, the sub-luminous 1991bg-like objects 
\citep{Filippenko:1992,Leibundgut:1993,Turatto:1996},
and the 1991T-like objects \citep{Filippenko:1992a,Phillips:1992}.
The 1991bg-like objects are photometrically conspicuous, having rapidly
evolving light-curves, and do not follow a simple linear Phillips
relation \citep{Phillips:1999}. Subsequently, several sub-classes
of SNe~Ia were invented based on spectroscopic features of the SNe
Ia and their velocities \citep{Benetti:2005,Branch:2009,Wang:2009}.
Of particular interest is what this variety of sub-classes tells us
about different progenitor models for SNe~Ia and whether the sub-classes
originate from distinct progenitor channels, or result from a single
channel with varying physical conditions %

In this paper, we investigate the photometric properties of SNe~Ia
using the CSP first and second data release \citep{Contreras:2010,Stritzinger:2011},
focusing on broad-band colors. In \S \ref{sec:CSPIdata} we briefly
describe the CSP photometry used for this analysis. We compare the
use of different light-curve parameters and their relation to other
photometric properties in \S \ref{sec:Decline-Rate-Parameter}.
In \S \ref{sec:Intrinsic-Colors} we examine the intrinsic colors
of SNe~Ia, revisit the Lira relation \citep{Lira:1998}, and examine the possible
reddening laws that could lead to the observed colors. The results
are summarized in \S \ref{sec:summary}.

\section{The CSP Data}\label{sec:CSPIdata}

In this section we briefly describe the CSP data set and the methods
used to extract photometric and spectroscopic parameters used in the
analysis. The CSP was proposed to be a next generation
low-redshift SN survey
building on the success of earlier surveys, yet providing a more well-defined
and calibrated photometric system. The large fraction of photometric
nights available at Las Campanas Observatory (LCO) has allowed us
to calibrate our photometry internally to a precision of 1\% 
\footnote{By internal precision, we mean that repeated observations of
local standards in the field of the SNe~Ia are consistent to 1\%.}. We have
also employed a monochrometer to accurately scan the entire
optical path of our telescopes at LCO, allowing for precise transformations
of our natural photometry to other systems \citep{Stritzinger:2011}.
Recent work by \citet{Mosher:2012}
showed that with such transformations, one could achieve 1\% agreement
in $ugri$ between the CSP and SDSS-II photometry.

For this paper, we utilize the first and second data release (DR1+DR2) sample of
the CSP described in \citet{Stritzinger:2011} and \citet{Contreras:2010},
augmented with additional
objects that will be published in the third data release (Krisciunas
et al., (in prep)). The sample of SNe~Ia used for this paper are listed
in Table \ref{tab:SNe}. The data reduction steps are fully outlined
in \citet{Contreras:2010}. The latest versions of our optical filter functions
are given in \citet{Stritzinger:2011} and are available on the CSP
website%
\footnote{\url{http://obs.carnegiescience.edu/CSP}%
}. 

When analyzing the photometric properties of the SN sample as a whole,
we simultaneously fit the light-curves of all available filters 
using SNooPy \citep{Burns:2010jx}.
SNooPy uses light-curve templates trained on a subsample of DR1+DR2 to
fit the decline rate parameter\footnote{Defined as the
change in $B$-band magnitude from peak to 15 days after peak in the
rest-frame of the SN \citep{Phillips:1993}.}, $\Delta m_{15}(B)$, time of
maximum in the $B$ band, and magnitude at maximum in each filter.
All colors from SNooPy fits are therefore \emph{pseudo-colors} of
the type $B_{max}-V_{max}$. Unless explicitly stated,
all colors in this paper should be interpreted as
pseudo-colors.  SNooPy also computes and applies the
necessary $K$ corrections based on the \citet{Hsiao:2007} SN~Ia spectral
energy distribution (SED) templates, updated to include the NIR. 

SNooPy can also model the effects of extinction on the shape of the
light-curves due to the change in effective wavelength of the filters. However
this effect is dependent on the amount of extinction and the shape of the
reddening curve, which are quantities we are attempting to infer. We therefore
do not apply this correction when fitting the data, but rather apply it
as part of our modeling (see \S \ref{sub:color-model}).

Where possible, 
the heliocentric
redshift of the host galaxy is used to compute the $K$ correction. For
SNe~Ia whose host is too faint to obtain a spectrum, the redshift is estimated
from the SN~Ia spectrum itself. 

In order to infer the extinction properties of dust contained in the host galaxy, it
is necessary to correct for any foreground extinction due to the Milky Way.
We adopt the Milky Way (MW) reddening estimates of \citet{Schlafly:2011}
and convert those to extinctions in each filter using the reddening
law of \citet[hereafter F99]{Fitzpatrick:1999},
adopting a total-to-selective absorption $R_{V}=3.1$. 

The CSP DR1+DR2 sample has 85 SNe~Ia. For this paper, we exclude 3 objects
whose light-curves are peculiar: SN~2004dt, SN~2006bt, and SN~2006ot.  Two
other objects (SN~2005ku and SN~2006lu) have particularly poor NIR photometry,
exhibiting large night-to-night variation in flux that is inconsistent with the errors
and is due to particularly poor signal-to-noise. 
Lastly, the $H$-band data for SN~2007hx was removed as
there is no evidence of SN flux in the subtracted images.  This leaves us with 82 SNe~Ia, 66
of which have NIR photometric coverage.

\section{Decline-Rate Parameter Revisited}\label{sec:Decline-Rate-Parameter}

\citet{Phillips:1993} first showed that the accuracy
of SNe~Ia distances could be greatly improved by correcting for an
empirical correlation between how fast the SN~Ia evolves and its peak
luminosity. Since then, there have been two parameters widely used
to quantify the rate of evolution of the light-curve. The decline-rate
parameter, $\Delta m_{15}(B)$, was first presented by \citet{Phillips:1993}.
The stretch, first introduced by \citet{Perlmutter:1999}
is simply a time stretching factor that maps the observed rest-frame
$B$ light-curve to a typical average template. More recently, stretch
has been used on the underlying SED model rather than the light-curve
to achieve the same effect \citep{Guy:2005,Guy:2007,Conley:2008}.
In either case, one is essentially measuring the rate of evolution
of the object, in such a way that intrinsically brighter objects have
broader light-curves. In this section we re-visit $\Delta m_{15}(B)$ as
a light-curve parameter and show that at the fast end of the SN~Ia
population, it does not provide a reliable measure of the relative rates of evolution
of the SNe~Ia.

\begin{figure}
\includegraphics[width=3.5in]{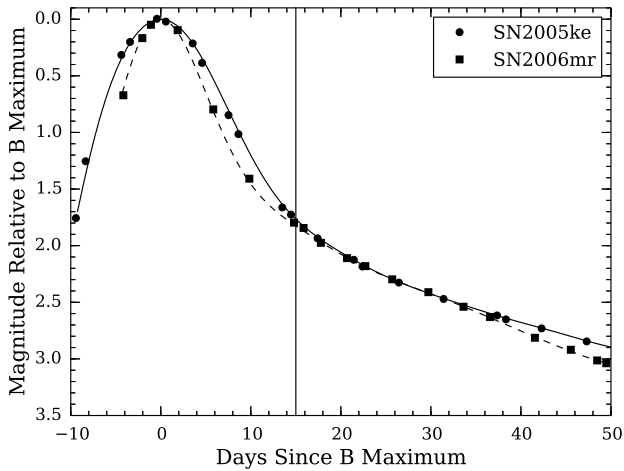}
\caption{Comparison of the rest-frame $B$ light-curves of SN~2005ke and SN~2006mr. The
measured decline rate $\Delta m_{15}(B)$ is nearly identical for
the two objects despite SN~2006mr having a faster rise time and decline
time prior to day 15. The solid and dashed lines are spline fits to
the data.\label{fig:SN2005ke+SN2006mr}}
\end{figure}

\subsection{The problem with $\Delta m_{15}(B)$}

With the increased number of SNe~Ia in our sample, particularly with
the faster-evolving events, we soon discovered that the $\Delta m_{15}(B)$
parameter had problems. Some of these problems are purely technical.
The first, identified early on by \citet{Leibundgut:1988} and \citet{Phillips:1999},
was that any reddening suffered by the SN~Ia would change the shape
of the $B$-band light-curve and therefore the observed value of $\Delta m_{15}(B)$
\footnote{Note that this is also a problem for a stretch derived using a $B$-band
template.}. A similar problem is that by definition, $\Delta m_{15}(B)$ is
tied to a particular photometric system, and so will vary from data set
to data set and would require S-corrections \citep{Suntzeff:1988,Stritzinger:2002} to
convert one set of $\Delta m_{15}(B)$ to another. And lastly, $\Delta m_{15}(B)$
is defined by measuring the light-curve at two very specific epochs,
and some form of interpolation is needed to measure these. All these
problems are mitigated somewhat by the use of light-curve template fitting
\citep{Hamuy:1996a,Prieto:2006,Jha:2007,Burns:2010jx}.

However, we have found that the very definition of $\Delta m_{15}(B)$
starts to break down as one approaches $\Delta m_{15}(B)\sim1.7$.
At this point, the change in intrinsic shape of the $B$ light-curve
becomes more complicated than a simple stretch relationship and the
decline in magnitude at 15 days after peak does not discriminate as
well between faster and slower evolving objects. An example of this
can be seen in Figure \ref{fig:SN2005ke+SN2006mr}, where the $B$-band
light-curves
of SN~2005ke and SN~2006mr are over-plotted. Clearly SN~2006mr has a
faster rise time and initially a faster decline after maximum, yet
their rest-frame light-curves happen to intersect near day 15.

While examining the $B-V$ light-curves, however, we noticed
that $\Delta m_{15}(B)$ failed to capture another trend in the photometric
behavior of fast SNe~Ia: the $B-V$ color as a function of time. In the
next section, we examine the $B-V$ light-curves of our sample and
show that the time of $B-V$ maximum may be a better parameter than
$\Delta m_{15}(B)$ for the fastest evolving SNe~Ia.

\begin{figure}
\includegraphics[width=3.5in]{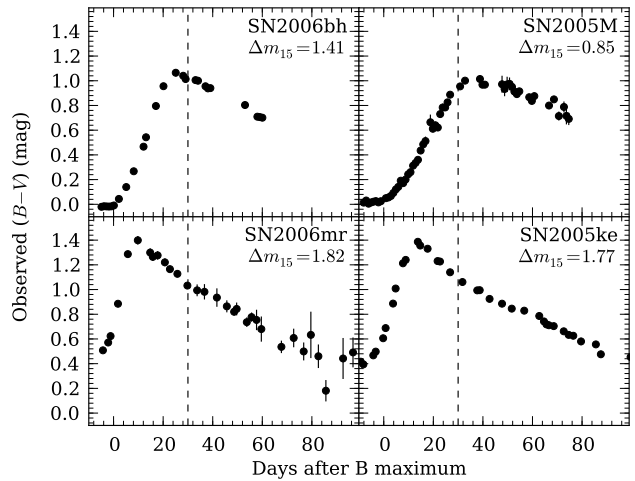}
\caption{Sample $B-V$ color-curves of 4 SNe~Ia, increasing in $\Delta m_{15}(B)$
from lower-left to upper-right. The time of $B-V$ maximum for the
fastest declining SN~Ia, SN~2006mr ($\Delta m_{15}(B)=1.82$), is near
10 days after $B$-maximum, whereas for SN~2005M ($\Delta m_{15}(B)=0.85$),
the maximum occurs approximately 40 days after $B$-maximum. A vertical
dashed line is plotted at $t=30$ days to more clearly show the progression
of peak times.\label{fig:BVsample}}
\end{figure}

\begin{figure}
\includegraphics[width=3.5in]{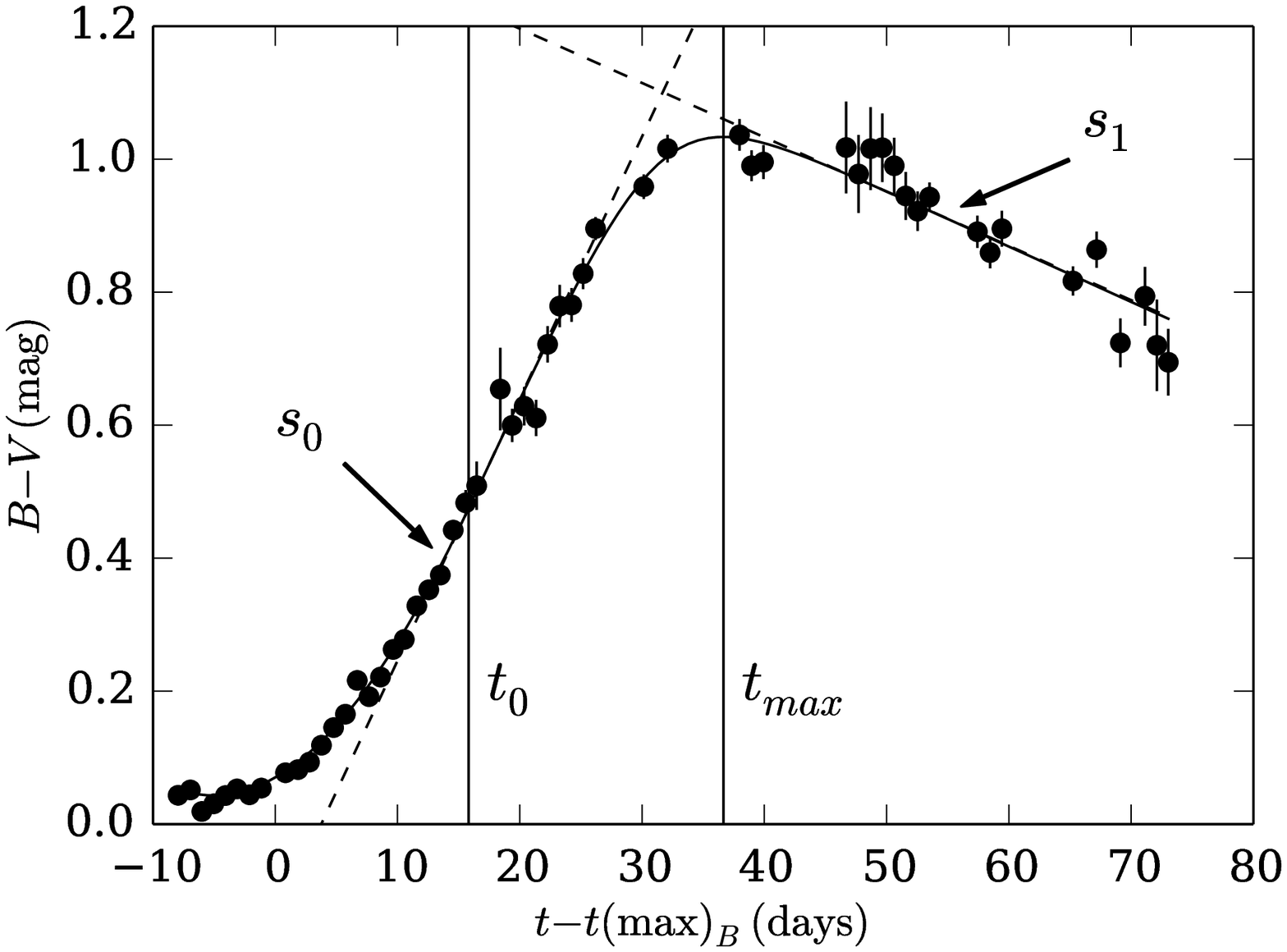}
\caption{Sample fit of the $B-V$ light-curve of SN~2005M using Equation
(\ref{eq:BVfit}). The points are the observed data, the solid line
is the best-fit model, the two dashed lines are the initial and final slopes $s_0$
and $s_1$,
and the two vertical lines denote $t_{0}$ and $t_{max}$.\label{fig:BVfit}}
\end{figure}

\subsection{The $B-V$ Color-Curve}

As one might expect, the shape of the $B-V$ color-curve has a more
complicated morphology than either the $B$ or $V$ broad-band light-curves. 
The
general shape of the color-curve is a local minimum (bluest) near the time
of maximum, followed by a near linear increase to a local maximum
(reddest) near day 30, followed by a linear decline out to later times.
\citet{Lira:1996} noticed that while the early time morphology of
$B-V$ varied with $\Delta m_{15}(B)$, the late-time linear decline
was remarkably consistent for those objects believed to have little or no interstellar
reddening. Use of this linear decline as a standard color is
termed the Lira relation (see \S \ref{sub:The-Lira-Law}).

As one examines the behavior of the $B-V$ color-curves, it is quickly
apparent that the location of the $B-V$ maximum is highly correlated
with $\Delta m_{15}(B)$. Figure \ref{fig:BVsample} shows 4 SNe~Ia
from our sample that have a range of decline rates. Clearly, the faster
objects have $B-V$ maxima at earlier times compared to more slowly
evolving objects.

\begin{figure*}
\includegraphics[width=3.5in]{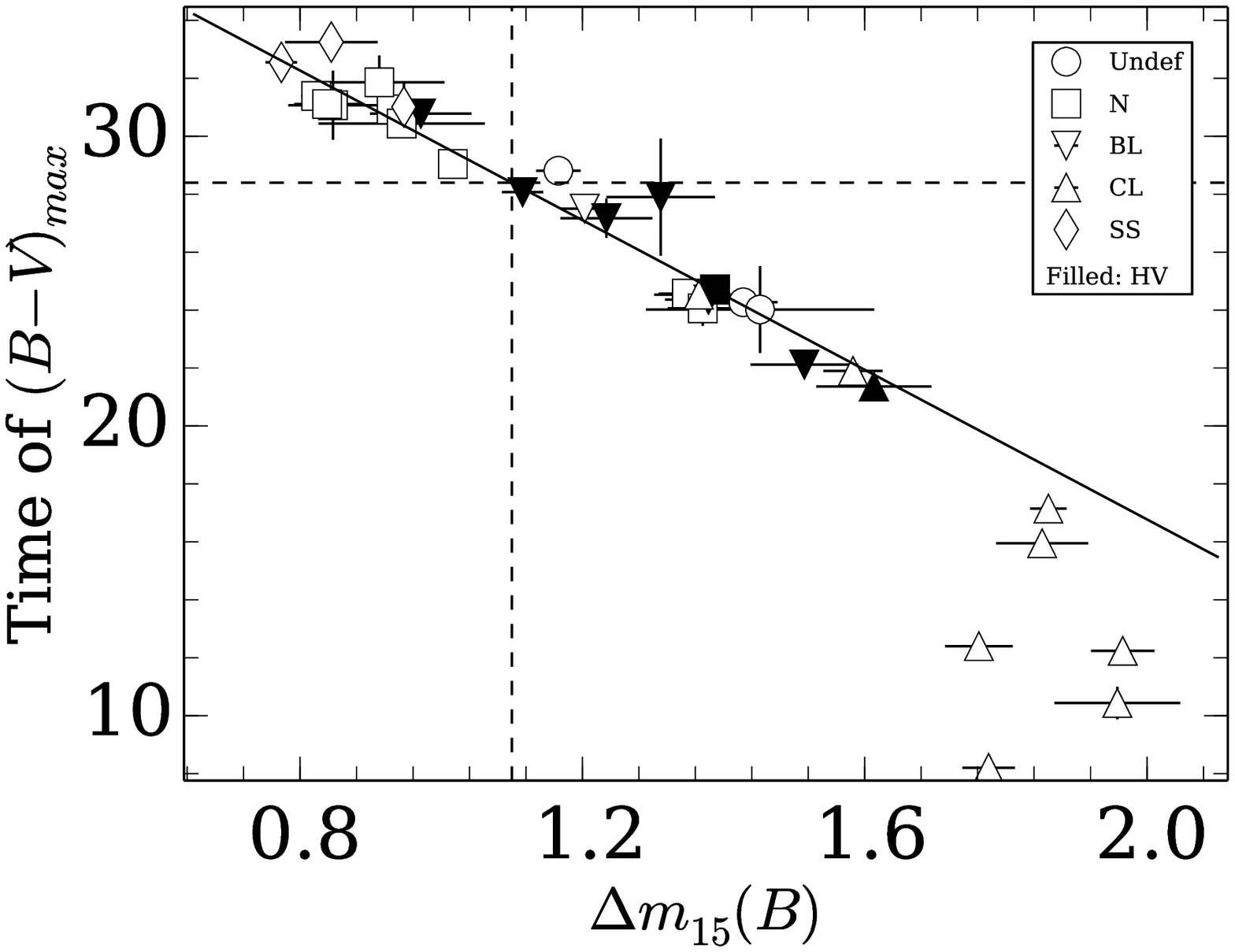}\includegraphics[width=3.5in]{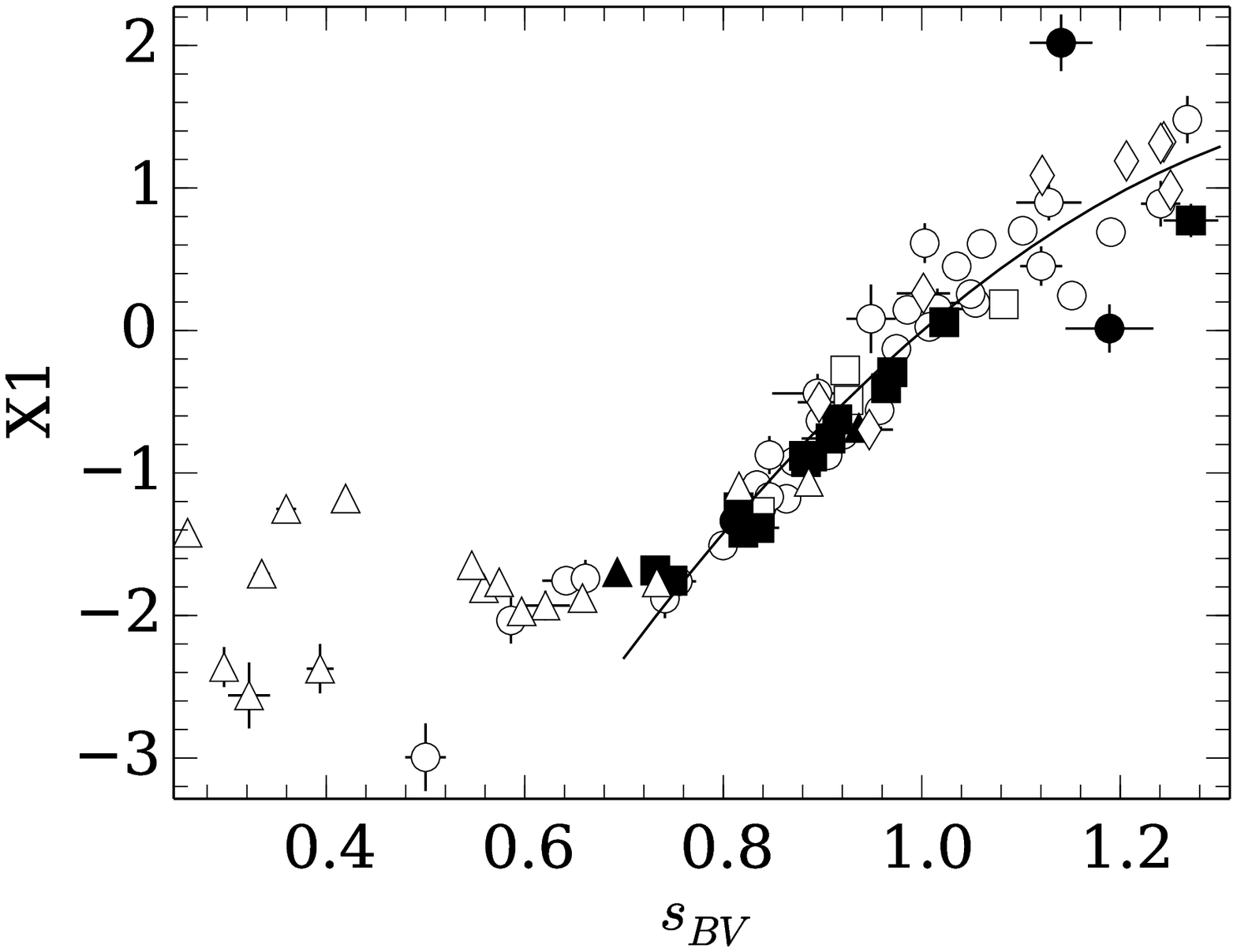}
\caption{(Left) Correlation between the rest-frame time of $B-V$ maximum, $t_{max}$ 
with observed
$\Delta m_{15}(B)$. Different plot symbols are used to indicate the
spectroscopic classification of \citet{Folatelli:2013} based on the
system of \cite{Branch:2009}:
squares are ``Core Normal'' (N), downward-pointing triangles are ``Broad
Line'' (BL), upward-pointing triangles are ``C ool'' (CL), diamonds
are ``Shallow Silicon'' (SS), and circles are unclassified. Filled
symbols are further classified as ``High Velocity''. There is a
very good correlation for $\Delta m_{15}(B)<1.7$, whereas the two
observables become decoupled for the faster decliners, which correspond
to the CL type. The solid line is the fit given by Equation 
(\ref{eq:BVmax_vs_dm15}).
The vertical dashed line shows $\Delta m_{15}(B)=1.1$ and the horizontal
dashed line shows the corresponding value of $t_{max}=28.65$. (Right) 
Relationship between $s_{BV}$ and the SALT $x_1$ parameter. The solid
line is the fit given by Equation (\ref{eq:x1_vs_sBV}).
\label{fig:BVmaxdm15}}
\end{figure*}

In order to investigate the $B-V$ behavior more quantitatively, we
require an analytic function that will give us an estimate of the
time of $B-V$ maximum as well as a fit to the late-time linear
decline. We therefore require a function that is nearly linear with
positive slope at early time, reaches a maximum, then transitions
to linear with negative slope. The derivative of the function can
therefore be
described by:
\begin{equation}
   y^{\prime}(t)=\frac{s_{0}+s_{1}}{2}+
   \left(\frac{s_{1}-s_{0}}{2}\right)\tanh\left(\frac{t-t_{max}}{\tau}\right),
   \label{eq:BVslope}
\end{equation}
where $s_{0}$ is the initial slope, $s_{1}$ is the final slope,
$t_{max}$ is the location of the maximum, and $\tau$ is the length
scale over which the transition occurs (the sharpness of the peak).
Integrating this will give the required function. One can also add
a polynomial term to capture the earliest behavior. The final function
can therefore be written as:
\begin{eqnarray}
   y(t) = &\frac{\left(s_{0}-s_{1}\right)}{2}+
   \frac{\tau}{2}\left(s_{1}-s_{0}\right)\ln\left[\cosh\left(\frac{t-t_{max}}{\tau}\right)\right]
   \nonumber \\
   & +c+f_{n}(t,t_{0}),
   \label{eq:BVfit}
\end{eqnarray}
where $c$ sets the overall normalization and $f_{n}(t,t_{0})$ is
an order $n$ polynomial for $t<t_{0}$ and equal to 0 for $t>t_{0}$.
Figure \ref{fig:BVfit} shows a sample fit to the $B-V$ color-curve
of SN~2005al using a quadratic to fit the early-time data. As can be
seen, the two parameters of most interest ($t_{max}$ and $s_{1}$)
are well-determined by the fit.

With best-fit values of $t_{max}$ from the $B-V$ color-curves,
we can investigate the correlation with the light-curve shape. 
The left-hand graph of Figure
\ref{fig:BVmaxdm15} shows this correlation. Indeed, for the lower
values of $\Delta m_{15}(B)$, there is a very strong correlation
with the time of $B-V$ maximum. However, for $\Delta m_{15}(B)>1.7$,
the correlation breaks down. 
We fit a straight line to the data with $\Delta m_{15}(B)<1.7$
and get the following relation between the two light-curve parameters:
\begin{equation}
t_{max}=28.65(13)-13.74(58)\left[\Delta m_{15}(B)-1.1\right].\label{eq:BVmax_vs_dm15}
\end{equation}

The question now arises: is $t_{max}$ a better light-curve parameter
than $\Delta m_{15}(B)$? For convenience, we now define
a dimensionless stretch-like parameter $s_{BV}=\frac{t_{max}}{30\ \mathrm{days}}$.
To differentiate from other parameters, we will call this the ``color-stretch''.
Together with Equation (\ref{eq:BVmax_vs_dm15}), we can convert between
$s_{BV}$ and $\Delta m_{15}(B)$ using the formula:
\begin{equation}
   s_{BV}=0.955-0.458\left(\Delta m_{15}(B)-1.1\right)\label{eq:sBV_vs_dm15},
\end{equation}
which is valid for $\Delta m_{15}(B)<1.7$. Likewise, we can compare
$s_{BV}$ to another commonly used light-curve shape parameter: the $x_{1}$
parameter in the SALT light-curve fitter \citep{Guy:2007}. The
right-hand graph of Figure
\ref{fig:BVmaxdm15} shows the relationship between the two parameters.
Clearly there is a strong correlation for $s_{BV}>0.7$.
Fitting a second order polynomial, we find a formula relating $s_{BV}$
to $x_{1}$ for $s_{BV} > 0.7$:
\begin{equation}
x_{1}=-0.006+5.98\left(s_{BV}-1\right)-5.55\left(s_{BV}-1\right)^{2}.\label{eq:x1_vs_sBV}
\end{equation}

From a technical standpoint, measuring $s_{BV}$ instead of $\Delta m_{15}(B)$
offers the advantage of being relatively insensitive to reddening. To demonstrate this,
we multiplied the \citet{Hsiao:2007} SED by F99 reddening curves with varying 
amounts of extinction and constructed synthetic $B-V$ color-curves. Measuring the
time of $B-V$ maximum, we found systematic shifts in $t_{max}$ 
of no more than 0.2 days for extinctions up to $A_V = 3$ mag. This corresponds to
changes in $s_{BV}$ of less than 1\%. A disadvantage of $s_{BV}$ is that to measure
it directly, one must have restframe $B$ and $V$ coverage from the time of $B$ maximum until
approximately 40 days. Here again, one must resort to template fitting
when such data are lacking.
In the next section we examine how well $s_{BV}$ captures the photometric
diversity of SNe~Ia . $ $

\begin{figure*}
\includegraphics[width=3.5in]{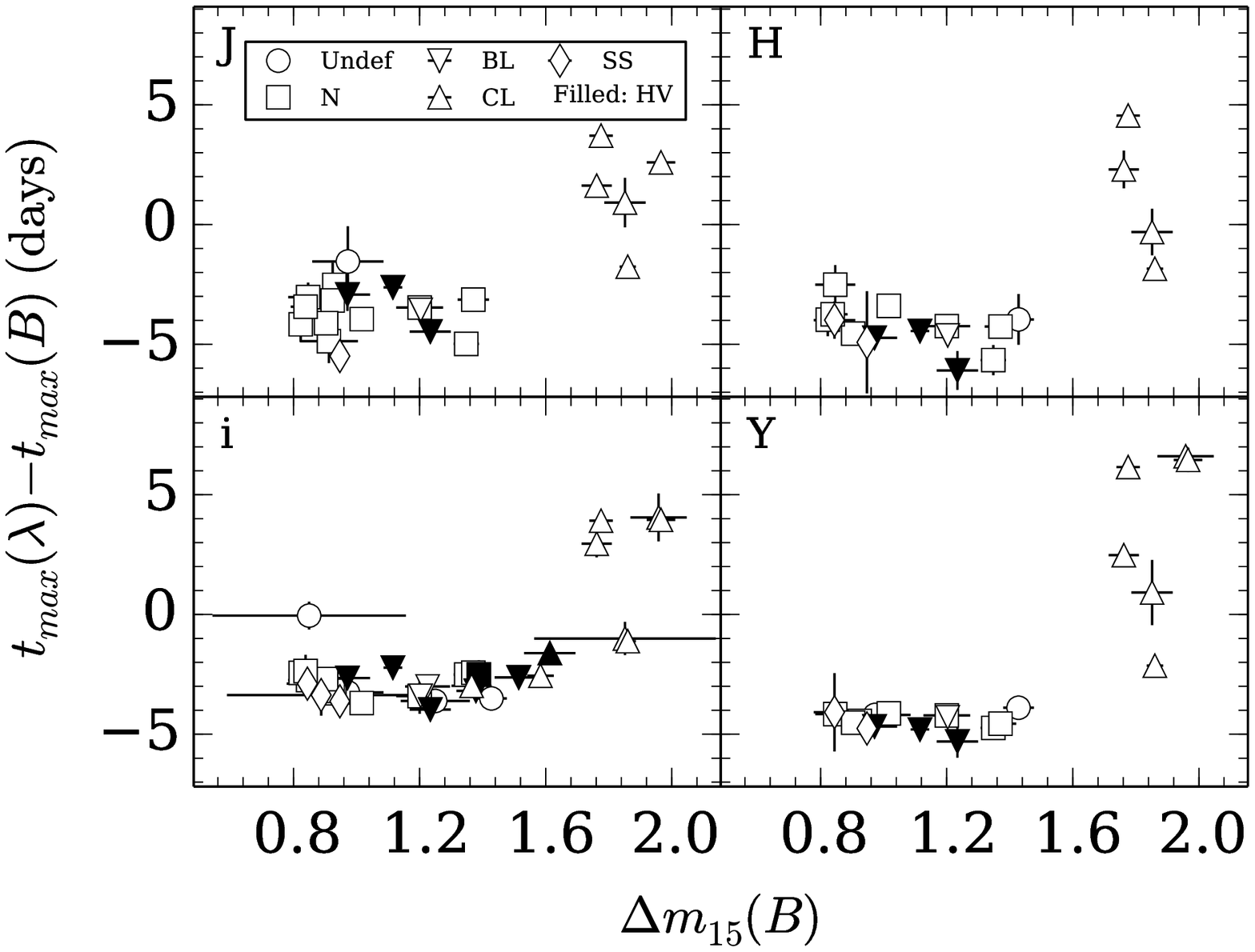}\includegraphics[width=3.5in]{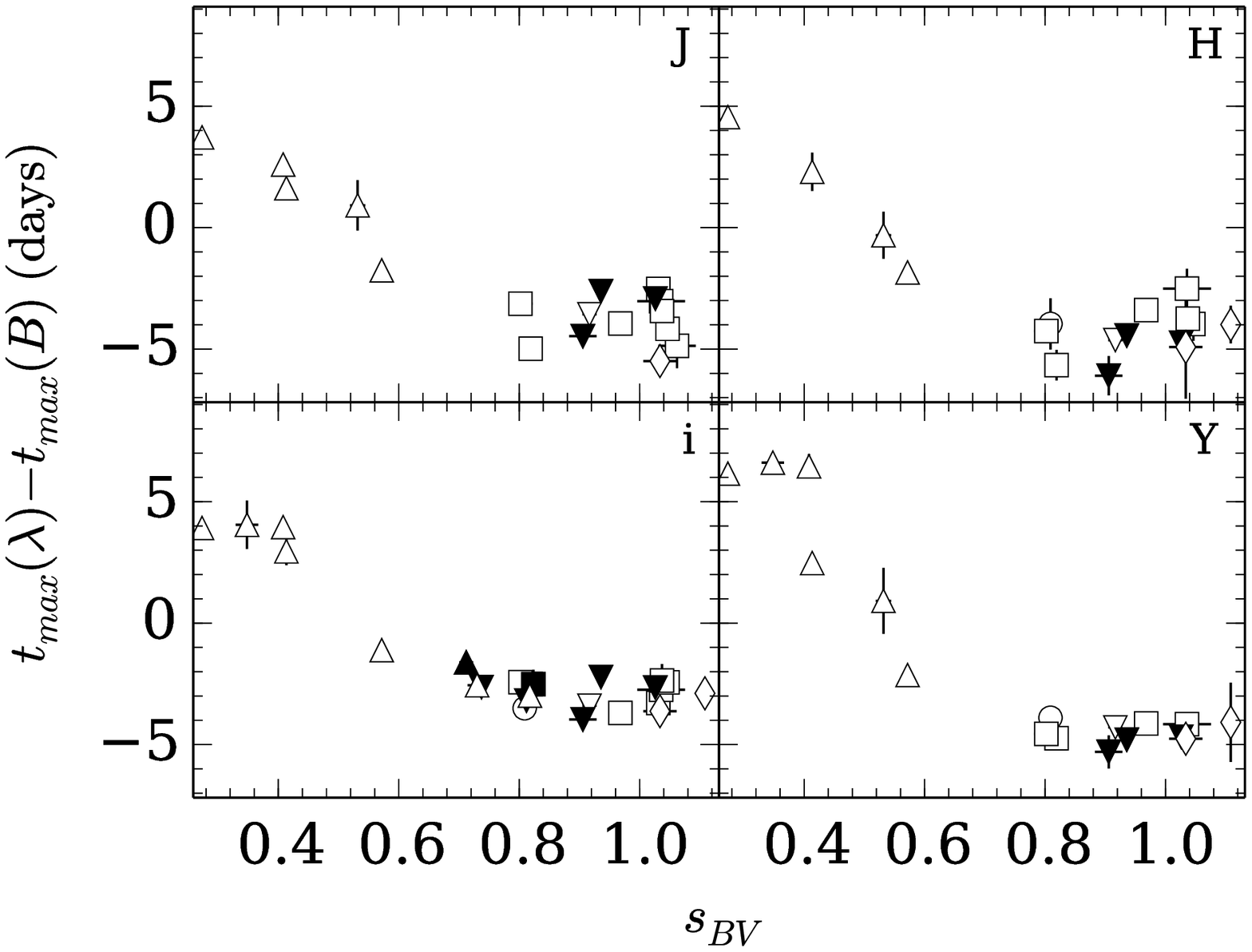}
\caption{The time of maximum in the NIR bands as a function of decline rate
$\Delta m_{15}(B)$ (left) and $B-V$ color-stretch $s_{BV}$ (right). Each panel represents
a separate filter in the set $iYJH$. The meaning of the plot symbols
is the same as Figure \ref{fig:BVmaxdm15}. \label{fig:NIR_Tmax}}
\end{figure*}

\subsection{NIR Light-Curves of Low Stretch SNe~Ia}

It has been known for some time that there is a dramatic change in
the morphology of SNe~Ia as one observes redward of the $r$ band 
\citep{Elias:1981,Ford:1993,Hamuy:1996b}. 
For ``normal'' SNe~Ia ($\Delta m_{15}(B)<1.7$), the primary maximum
is followed by a secondary maximum approximately 20 days later. However,
as one moves to higher $\Delta m_{15}(B)$, the secondary maximum
disappears and we are left with a single peak. The question then arises:
is this a continuous diminishing of the strength of the secondary
peak as $\Delta m_{15}(B)$ increases (or as color-stretch decreases), or
are we dealing with two separate sub-populations of SNe~Ia \citep{Krisciunas:2009}?
While it is not possible to answer this question definitively based
on photometric data alone, finding a ``missing link'' between the
slow decliners (with clear prominent NIR secondary peak) and the fast
decliners (with no secondary NIR peak) would certainly lend credence
to the former scenario. We therefore turn to examining the properties
of the NIR light-curves as a function of how fast they evolve.

\subsubsection{The Time of NIR Maxima}

The left-hand graph of Figure \ref{fig:NIR_Tmax} shows the time of 
first maximum of the NIR light-curves
as a function of the decline rate $\Delta m_{15}(B)$. 
In general, slowly declining SNe~Ia peak in the NIR 2-6 days earlier
than $B$ band, which presents an observational challenge to obtaining
NIR coverage of the peak for objects whose observations are typically
triggered based on optical band discoveries. However, it has been known for some time
that the faster evolving events tend to peak later (after $B$-band
maximum). Examining the right-hand graph in Figure \ref{fig:NIR_Tmax},
it would appear that
we are seeing two groups of objects: a set with early NIR peak and
a set with late NIR peak, which has been seen before
\citep{Krisciunas:2009,Phillips:2012}. The early NIR peak objects span the ``normal''
range $0.7<\Delta m_{15}(B)<1.7$, while the late NIR peak objects are
exclusively in the fast-declining region $\Delta m_{15}(B)>1.7$. And
in the fast-declining region, there is a large spread of peak times
with no correlation with $\Delta m_{15}(B)$. On the other hand, when
examining the right-hand graph of  Figure \ref{fig:NIR_Tmax},
where we have used the $s_{BV}$
parameter, the time of NIR peak is clearly
correlated with $s_{BV}$.
As a result, it can be argued that the bifurcation of the fast-declining
sample into early- and late-risers is an artifact of the 
behavior of the light-curve parameter $\Delta m_{15}(B)$.

\subsubsection{The Strength of the 2nd Peak}

\begin{figure}
\includegraphics[width=3.5in]{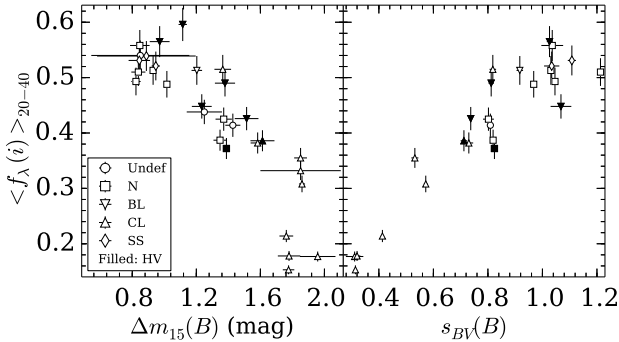}
\caption{The average normalized flux of the 
$i$-band between days 20 and 40
in the rest-frame of the SN as a function of the decline rate 
$\Delta m_{15}(B)$ (left panel)
and the color-stretch $s_{BV}$ (right panel). 
The meaning of the plot symbols
is the same as in Figure \ref{fig:BVmaxdm15}.\label{fig:i-flux}}
\end{figure}

A second conspicuous feature of the NIR light-curves for normal SNe~Ia 
is the presence of a secondary maximum. This is due to recombination
of iron group elements in the supernova ejecta \citep{Kasen:2006}
and begins to be seen in $r$ band as an inflection point, and develops
into a secondary maximum in $i$ band. In the $z$ band and $Y$ band, the secondary
maximum can in fact be stronger than the primary 
\citep{Stritzinger:2002,Burns:2010jx}.
It becomes less
prominent in $J$ band, and then regains strength in the $H$ band.
For the fast-declining SNe~Ia, however, both optical and NIR light-curves
show either a very weak secondary or none at all. Following \citet{Krisciunas:2001},
we consider the strength of the $i$ band secondary maximum as a function
of decline rate. They define $\left\langle f_{\lambda}(i)\right\rangle _{20-40}$
as the average flux (normalized to maximum) in the $i$-band from day 20 to day
40 after $B$ maximum in the rest-frame of the SN~Ia. This gives a
measure of the prominence of the secondary peak. In the left panel
of Figure \ref{fig:i-flux}
we plot this as a function of $\Delta m_{15}(B)$. We
see a clear correlation, albeit with significant scatter
(see also Figure 8 of F10),
and the fast decliners seem to separate into two groups. However,
if we instead plot against $s_{BV}$ (see right panel of Figure \ref{fig:i-flux}),
there is again a more continuous transition between the slow and fast
decliners.

\subsubsection{Continuity of Light-Curve Templates}

The transition (if one exists) from the slow to fast declining SNe~Ia
has been a major stumbling block for light-curve fitters. Clearly,
a simple stretch in time cannot account for the NIR light-curves losing
the secondary peak as we go from slow decliners to fast decliners.
To our knowledge, SNooPy is the only light-curve fitter that currently
attempts to fit NIR light-curves for both slow and fast decliners.
Up until recently, SNooPy's NIR templates for the fast-decliners were
unreliable due to two problems: 1) small numbers of fast-declining
SNe~Ia with well-observed light-curves, and 2) the apparent failure
of the $\Delta m_{15}(B)$ parameter to distinguish between the fastest
decliners.

\begin{figure}
\includegraphics[width=3.5in]{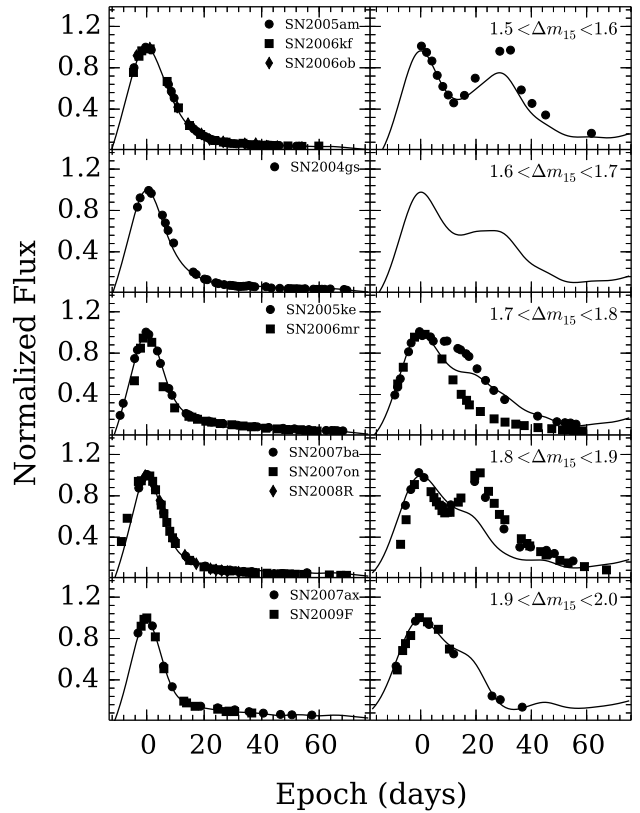}
\caption{Light-curves of 11 SNe~Ia for which $1.5<\Delta m_{15}(B)<2.0$, binned
into 5 intervals. The left panels show $B$ band and right panels show $Y$ band.
The individual SNe~Ia are labeled with different symbols. The average
SNooPy light-curve template for the bin is shown as a black line.
Note that one panel is missing data due to the fact that the
object SN~2004gs has no NIR observations.\label{fig:fast_dm15}}
\end{figure}

\begin{figure}
\includegraphics[width=3.5in]{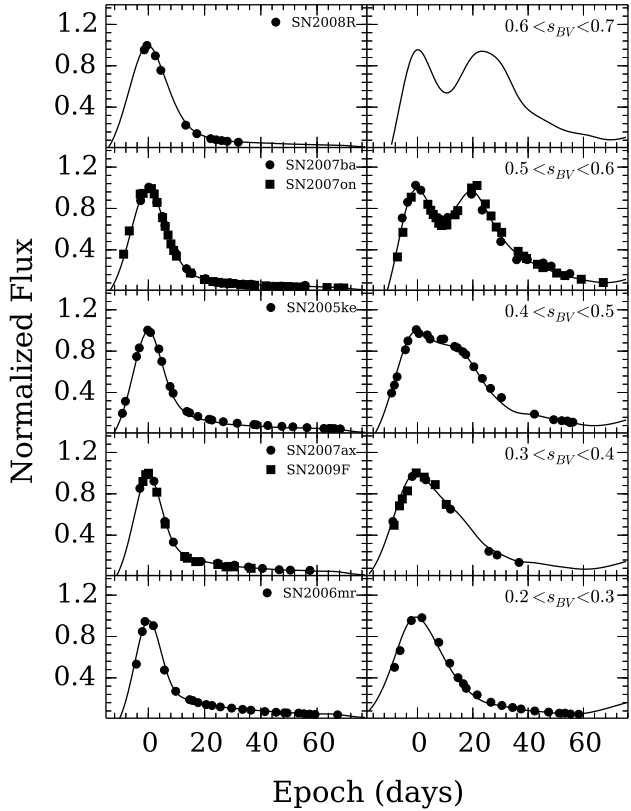}
\caption{Same as Figure \ref{fig:fast_dm15}, but using the color-stretch parameter,
$s_{BV}$, to bin the SNe~Ia having $0.2<s_{BV}<0.7$. Note that one panel is
missing data due to the fact that object SN~2008R has no NIR observations.
\label{fig:fast_BVmax}}
\end{figure}

Figure \ref{fig:fast_dm15} shows the sorting of the 11 fastest decliners
in the SNooPy training sample into bins of $\Delta m_{15}(B)$. One
can clearly see the problem: while the $B$-band light-curves seem
to be sorted out correctly (with the possible exception of SN~2006mr), the $Y$-band
light-curves clearly do not follow any kind of continuous progression
from $\Delta m_{15}(B)\simeq1.5$ to $\Delta m_{15}(B)\simeq2.0$.
The same holds true for the other NIR bands. The situation, however,
is completely different when one uses the $s_{BV}$ parameter instead.
In Figure \ref{fig:fast_BVmax}, we can see that SN~2006mr, which is
clearly the fastest declining of the SNe~Ia, is now the lowest color-stretch. SN~2007ba
and SN~2007on are now categorized as slower events. This has the very
noticeable effect that the $Y$-band light-curves now show a more
continuous progression from $s_{BV}\simeq0.2$ to $s_{BV}\simeq0.6$.
Another notable feature is that SN~2005ke now looks very much like
a transition object, having no secondary peak, but clearly an inflection
point. 

\begin{figure}
\includegraphics[width=3.5in]{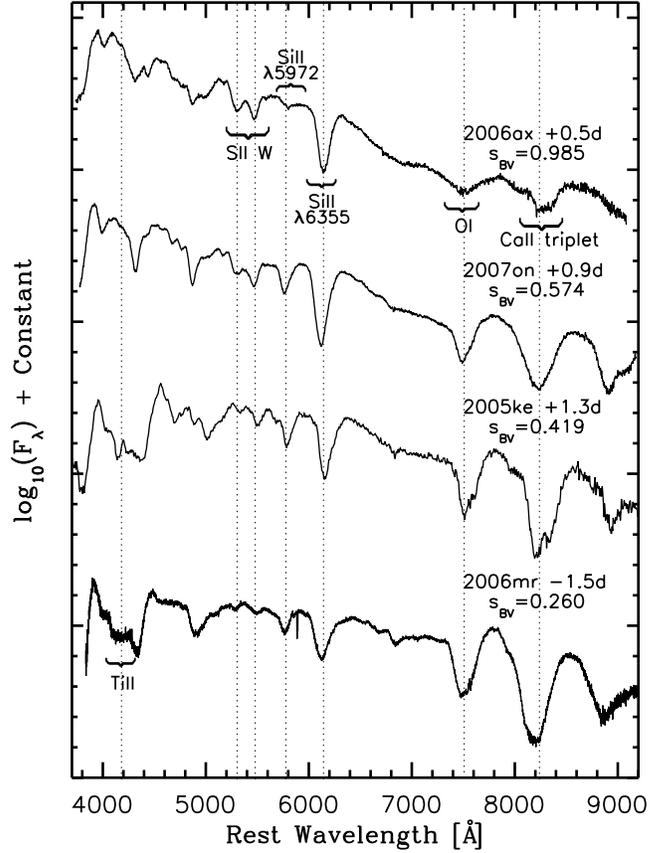}
\caption{Comparison of the spectra of 4 SNe~Ia near maximum light. The spectra are
labeled with the object's name, phase, and color-stretch $s_{BV}$. Several prominent
spectral features are also labeled:
\ion{Ti}{2}~$\lambda 4250$, the \ion{S}{2} ``W'' feature~$\lambda\lambda 5454,5640$,
\ion{Si}{2}~$\lambda 5972$, \ion{Si}{2}~$\lambda 6355$, 
\ion{O}{1}~$\lambda\lambda 7772, 7775$, and the \ion{Ca}{2} IR triplet $\lambda 8579$.
Note that due to high expansion velocities,
these lines will appear to the blueshifted relative to their rest wavelengths in
observed spectra.\label{fig:spec_comp}}
\end{figure}

\begin{figure}
\includegraphics[width=3.5in]{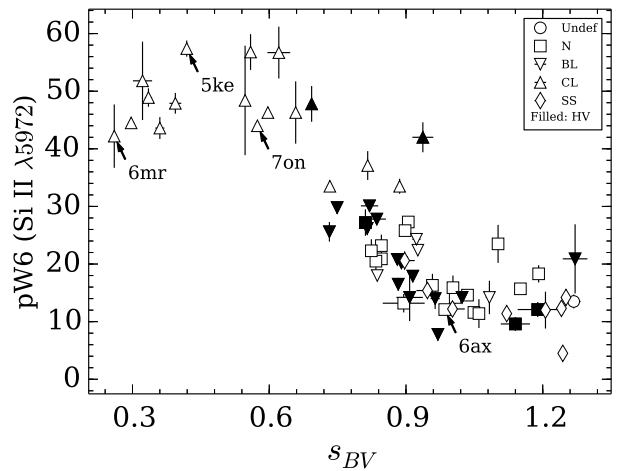}
\caption{The pseudo-equivalent width of the \ion{Si}{2}~$\lambda 5972$ feature as a function
of the color-stretch $s_{BV}$.  The four objects from Figure \ref{fig:spec_comp} are
labeled. The meaning of the plot symbols
is the same as in Figure \ref{fig:BVmaxdm15}.\label{fig:sBV_pw6}}
\end{figure}

To further investigate this, we examined the spectra of 4 representative
objects in this range of $s_{BV}$: SN~2006ax ($s_{BV} = 0.985$), SN~2007on
($s_{BV}=0.574$), SN~2005ke ($s_{BV}=0.419$), and SN~2006mr ($s_{BV}=0.260$).
Figure \ref{fig:spec_comp} shows the spectra of these 4 objects near maximum
light.
SN~2005ke shows several spectroscopic features common to the ``Cool''
SNe~Ia: prominent \ion{Ti}{2} and \ion{Si}{2}~$\lambda 5972$. On the other
hand, it shows an intermediate strength in the \ion{S}{2} ``W'' feature.
The presence of \ion{Ti}{2} is highly sensitive to effective temperature and 
will therefore turn on rather abruptly, making any transition with
$s_{BV}$ equally abrupt. Another feature know to be correlated with
light-curve width is \ion{Si}{2}~$\lambda 5972$ \citep{Folatelli:2013}.
In Figure \ref{fig:sBV_pw6} we plot the pseudo-equivalent width of
\ion{Si}{2}~$\lambda 5972$ from \citet{Folatelli:2013} as a function of
$s_{BV}$. Interestingly, there is a more complicated relation than the 
linear trend seen with $\Delta m_{15}$. In particular, there seems to
be very little trend for $s_{BV} < 0.5$. 

While it remains unclear whether $s_{BV}$ is in fact a ``better'' parameter
than $\Delta m_{15}$ in predicting the physics of the SNe~Ia,
the fact remains that it does produce a much more continuous family of light-curve
templates for the faster-declining objects. For this reason, the
SN~Ia template-fitting package SNooPy will incorporate templates parametrized by
$s_{BV}$ and we will adopt it as our shape parameter for the remainder of
the paper. It is also unclear whether $s_{BV}$ will present a superior alternative
when analysing high-redshift SNe~Ia for the purposes of cosmology, as the 
fast-declining events are less numerous and significantly fainter than 
normal SNe~Ia.

\section{Intrinsic Colors of SNe~Ia}\label{sec:Intrinsic-Colors}

As shown by numerous groups 
\citep[e.g.,][hereafter F10]{Tripp:1998,Phillips:1999,Astier:2006,Folatelli:2010},
in order to correct for the interstellar reddening and any other intrinsic
color-luminosity relationship for the purposes of determining distances,
it suffices to use the observed colors. However, if we wish to understand
what is causing the observed color distribution of SNe~Ia or wish
to measure their intrinsic luminosities, then we must determine the
intrinsic colors of these objects in order to separate out the contribution
from dust extinction along the line of sight. In the following sections,
we present two methods for determining these intrinsic colors: the
Lira relation and statistical inference from the observed
pseudo-colors.

\subsection{\label{sub:The-Lira-Law}The Lira Relation Revisited}

\citet{Lira:1996} discovered that the $B-V$ color-curves of SNe~Ia,
while showing significant variation at early times, all seem to converge
to a linear decline at late times. By fitting a line to the late-time
$B-V$ data, one could then determine the amount of reddening relative
to an ensemble of objects for which it is assumed (see \S \ref{sec:low-red})
there is little
to no reddening \citep{Phillips:1999}. As far as could be seen with
the data at the time, there did not appear to be any significant difference
in the late-time colors of either slow or fast decliners, providing
a truly standard color. With our improved number of objects and homogeneous
photometry, we re-visit the Lira relation, as was done in F10.

\begin{figure}
\includegraphics[width=3.5in]{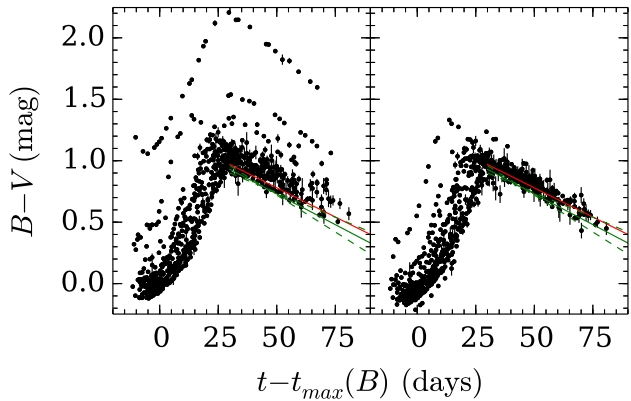}
\caption{The $B-V$ color-curves for 40 SNe~Ia in our sample. 
The left panel shows $B-V$ corrected for Milky-Way extinction
only, whereas the right panel shows $B-V$ curves shifted downward so
as to minimize the residuals with respect to the
Lira relation from F10, shown as a solid red line in both panels.
The new fit given by Equation (\ref{eq:lira_updated}) for $s_{BV}=1$
is plotted as a solid green line and the observed range of slopes
are plotted as dashed green lines.
\label{fig:lira_BV}}
\end{figure}

Figure \ref{fig:lira_BV} shows the $B-V$ light-curves of 40 of our
SNe~Ia, for which reliable times of $B$-maximum are measured and
sufficient late-time ($t-t_{max}>30$ days) $B$ and $V$
photometry are observed
to fit the Lira relation. In the left-hand panel, we plot the rest-frame
\footnote{By rest-frame, we mean that the object's phase has been corrected
for time dilation and its observed flux has been $K$ corrected.}
$B-V$ color-curves corrected
for MW reddening derived from \citet{Schlafly:2011}.
In the right-hand panel, we offset each light-curve downward such that the
resulting late-time data show a minimum dispersion with respect to the
Lira relation from F10, which amounts
to 0.04 mag rms. In both panels we plot the Lira relation derived
by F10 in red. It is apparent that there is a systematic
difference between the mean slope of the corrected data and the F10
fit, which was done using a subset of the sample of SNe~Ia that were
believed to have suffered little or no reddening. In order to determine the reason
for the difference in slope, we examine each object individually.

\begin{figure}
\includegraphics[width=3.5in]{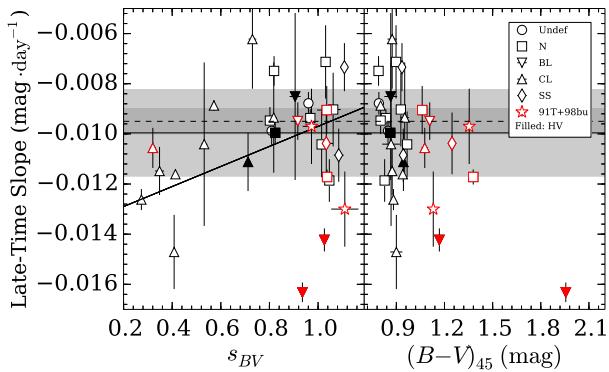}
\caption{The slope of the late-time $B-V$ color-curve as a function of 
the color-stretch, $s_{BV}$ (left panel), and the $B-V$ color at day 45, 
$(B-V)(45)$ (right panel).
The meaning of the symbols is the same as in Figure \ref{fig:BVmaxdm15}, however
objects whose $B-V$ at day 45 is greater than 1.0 mag are labeled
in red and two additional objects (SN~1991T and SN~1998bu) are labeled with stars.
The solid line in the left-hand panel is a fit to the data excluding the red
points. The horizontal dashed line and dark grey shaded region shows the
slope from F10 and the horizontal solid line and lighter shaded region show
the median and standard deviation of the slopes from this paper.
\label{fig:lira-slope}}
\end{figure}

For this purpose, we employ Equation (\ref{eq:BVfit}) to examine the
slopes ($s_{1}$) of the late-time linear portion of the $B-V$ light-curves
for our sample. Aside from getting an improved estimate of a global
Lira relation, we can examine in detail whether there are any trends
with respect to light-curve shape, reddening, and any other properties
we desire.

Figure \ref{fig:lira-slope} shows the late-time slope of the $B-V$
color-curves as a function of $s_{BV}$ (left panel) and $B-V$ color at day
45 (right panel). There appears to be a trend such that the fastest evolving objects
tend to have late-time $B-V$ slopes that are steeper (more negative)
than ``normal'' objects. This is to be expected if low-color-stretch objects
simply evolve more quickly than high-color-stretch objects. We note, however,
that the trend is not nearly as strong as the $1/s_{BV}$ trend one
would expect. We also see that there are significant outliers with
$s_{BV}\simeq1$ and very steep slopes. These also correspond to objects
with larger than average colors and we have labeled points for which
$(B-V)(45)>1$ in red. 
Objects that have redder colors tend
to have steeper late-time slopes. 
Looking at the most extreme case, SN~2006X, it is likely that the steepness
of the late-time slope is due a light-echo caused by scattering of
light by dust into our line of sight \citep{Wang:2008}. It is therefore
possible that the more moderately reddened cases also have scattering
of light into the line-of-sight, only to a lesser degree. 
Two other objects are known to have light-echoes: SN~1991T \citep{Schmidt:1994}
and SN~1998bu \citep{Cappellaro:2001}.
We plot them in Figure \ref{fig:lira-slope} as star symbols. SN~1998bu 
has a late-time slope consistent with most in our sample, whereas SN~1991T
follows the trend of the redder objects. While the light-echo in SN~2006X
is observed one month after maximum, those in SN~1991T and SN~1998bu
are detected much later (600 and 500 days after $B$ maximum, respectively).
It is possible that SN~1991T, like SN~2006X, had a light-echo at earlier
times that was missed due to the pecularity of this object. Then again,
being a peculiar SN~Ia, the steeper late-time $B-V$ slope could be
due to the physics of the object. It has been argued that this
change in slope could be due to CSM around these objects \citep{Forster:2013},
which is supported by the fact that both SN~2006X and SN~2007le are
known to have CSM and show very low $s_1$.

As it is not entirely clear whether or not
the trend of slope with color is intrinsic to the SN~Ia, we discard the 
red objects and fit a
linear relation between the late-time slope and $s_{BV}$, obtaining
$ds_1/ds_{BV} = 0.004 \pm 0.001$, which is shown in 
the left panel of Figure \ref{fig:lira-slope} as a solid line. We also solve
for an intrinsic $(B-V)(45)$ color and scatter by modeling the observed 
distribution as the convolution of a Gaussian with an exponential distribution
with scale length $\tau_{BV}$ \citep[see][]{Jha:2007}. We obtain 
$(B-V)(45) = 0.78 \pm 0.04$ and $\tau_{BV} = 0.19 \pm 0.03$. Together,
these results give us an updated Lira relation:
\begin{eqnarray}
(B-V)(t) & = & 0.78(04) - \left[ 0.0097(5) - \right.\nonumber \\
   & & \left.0.004(1)\left(s_{BV}-1\right)\right]
   \left( t - t_{max} - 45\right). \label{eq:lira_updated}
\end{eqnarray}
Indeed, the slope one obtains for a $s_{BV}=1$ object is quite consistent with
the slope derived in F10, shown as a dashed horizontal line and dark
shaded region in Figure \ref{fig:lira-slope}. The green solid line
in Figure \ref{eq:lira_updated} represents Equation (\ref{eq:lira_updated})
for $s_{BV}=1$ whereas the dashed green lines are for $s_{BV}=0.2$
and $s_{BV}=1.2$. 

\subsection{Low-Reddening Sample vs. Blue Edge}
\label{sec:low-red}
In order to derive the Lira relation, \citet{Lira:1996} used a ``low
reddening'' sample of SNe~Ia to anchor the colors to a standard intrinsic
locus. These objects were thought to have very little dust due to their distance from
the host and/or the host being an early type galaxy, which
are thought to have very little interstellar gas and dust \citep{Sternberg:2011}.
Later work by \citet{Phillips:1999} and F10
used similar arguments to construct a set of objects whose colors
were believed to be intrinsic, though also adding the requirement
that \ion{Na}{1} D absorption be absent from the objects' spectra. 
While this is a simple way to throw
out reddened objects for the purpose of determining intrinsic colors,
it also eliminates objects that have a small projected
impact parameter in a late type host which, if in the foreground,
could have no interstellar reddening. These objects also provide information
and we would like to keep them in the ``low reddening sample''. 

Furthermore, using host type and proximity as arguments for inclusion
in the reddening-free sample is predicated on the assumption that
the extinction is caused by the interstellar medium of the host,
whereas some portion of the extinction could be caused by dust associated
with the SN~Ia progenitor system itself. For this reason, we would
like to refrain from relying on the identification of a reddening-free
sample.

To this end, instead of categorizing the objects, we can exploit the
fact that extinction can only make objects redder. We therefore expect
that the bluest objects will define the intrinsic colors and therefore
we seek a ``blue edge'' to the data. From a Bayesian perspective,
we could simply apply a prior that the $B-V$ color excess,
$E(B-V)$, be strictly positive.
However this does not suffice, as this does not preclude arbitrarily
large positive values of $E(B-V)$ and arbitrarily large negative intrinsic
colors. We therefore must also insist that the likelihood that an
object is reddened decrease for large values of $E(B-V)$. The prior
will therefore be peaked and we address the functional form of the
prior in the next section.

\subsection{Inferring Intrinsic Colors\label{sub:color-model}}

Under the assumption that the extinction suffered by the SN follows
a well-defined dust law that is not uniquely grey, the amount of reddening
in each filter pair will be different. The most commonly used parameterization
has two quantities: $E(B-V)$ and the ratio
of total-to-selective absorption in the $V$ band, $R_{V}$. The color
excess sets the optical depth, while $R_{V}$ is related to the distribution
of dust grain sizes \citep{Weingartner:2001}. We further assume that
the intrinsic color of each SN~Ia is a smooth function of color-stretch
$s_{BV}$ and model this relationship as an $N$th order polynomial in
$P_j^N$. The observed colors are then given by 
\begin{eqnarray}
\left(B-m_{j}\right) &=& P_{j}^{N}\left(s_{BV}-1\right)+
\Delta A_{j}\left(E(B-V)_{host},R_{V}\right)+ \nonumber \\
 & & \Delta A_{j}\left(E(B-V)_{MW},3.1\right),
\label{eq:color_model}
\end{eqnarray}
where $m_{j}$ is the observed
rest-frame magnitude in filter $j=\left\{ u,g,r,i,V,Y,J,H\right\} $,
$\Delta A_{j}=A_{B}-A_{j}$ is the differential extinction between
$B$-band and filter $j$, being a function of the host galaxy color
excess $E(B-V)_{host}$, the host galaxy reddening parameter $R_V$,
and the MW foreground color excess $E(B-V)_{MW}$. 
$\Delta A_{j}$ is determined numerically by multiplying the \citet{Hsiao:2007}
SED with the appropriate reddening curve and computing synthetic 
photometry using the CSP filter functions.
We assume
the canonical MW value $R_V = 3.1$.
The form of the function $\Delta A_{j}\left(\right)$
is dictated by the choice of reddening law. For this paper, we consider
the reddening laws of \citet{Cardelli:1989,ODonnell:1994} (hereafter
CCM+O),
F99, and the CSM-motivated reddening
law of \citet[hereafter G08]{Goobar:2008}. For each color, we also
solve for an intrinsic scatter $\sigma_{xy}$. \citet{Chotard:2011}
and \citet{Scolnic:2014}
have shown the importance of accounting for intrinsic color variations and it
is our intent to do so with these parameters.

It is important to realize that the CCM+O and F99 dust laws
are empirical fits to observations of stars in our own MW's
interstellar medium (ISM). These fits were made to observations of
stars whose lines of sight have values of $R_{V}$ that cover the
range $2.6\le R_{V}\le5.6$ \citep{Cardelli:1989,Fitzpatrick:1999}. Therefore, allowing $R_{V}$ to vary
beyond these limits constitutes an extrapolation of these fits, which
of course is always dangerous. As we will see, several of our fits favor
$R_{V}<2$, where the dust laws are not valid. Nevertheless, the parameterization
of CCM+O, being of the form $a+b/R_{V}$, is very smooth and does
remarkably well at reproducing the observed colors of the
reddest SNe~Ia. We therefore
allow $R_{V}$ to vary below $R_{V}=2.6$, but refrain from inferring
anything about the physics of the dust or ISM.

For $N_{F}$ filters, we have $N_{F}-1$ independent colors. For each
color, we will fit the coefficients of $P_{j}^{N}$. For each supernova,
we will compute a separate reddening $E(B-V)_{host}$. The remaining parameter
of interest, $R_{V}$ can either be solved as a universal value for
the whole sample of SNe~Ia, or we can try to fit a separate $R_{V}$
for each object. We do both in the next sections. We also investigate
whether the polynomial $P_{j}^{N}$ should be a linear fit ($N=1,$
as was done in F10), or if the data warrant a quadratic ($N=2$).

Due to the multiplicative nature of $R_\lambda$ and $E(B-V)_{host}$ in 
the $\Delta A_{j}$ term
of Equation (\ref{eq:color_model}), these parameters will be highly
covariant. Also, as the $E(B-V)_{host}$ becomes small, the model becomes
less sensitive to $R_{V}$. We therefore require priors on these values.
For $E(B-V)_{host}$, we investigate three priors: 1) assigning zero reddening
to the ``low reddening sample'' (LRS) and uniform priors for every other
object; 2) an exponential prior for all objects as used by \citet{Jha:2007};
and 3) a truncated Cauchy distribution, which has a longer tail to
large values. For $R_{V}$, we investigate four priors: 1) a universal
$R_{V}$ for the entire sample with uniform prior; 2) separate $R_{V}$
for each SN~Ia with uniform priors; 3) separate $R_{V}$ drawn from
an N-component Gaussian mixture model; and 4) separate $R_{V}$ drawn
from one of $N$ independent Gaussians binned by $E(B-V)_{host}$. The last
of these is motivated by the results of \citet{Mandel:2011} who show a
linear trend between $R_V$ and $A_V^{-1}$. These priors
are discussed in more detail in Appendix \ref{sec:Appendix1}.

\begin{figure*}
\includegraphics[width=6in]{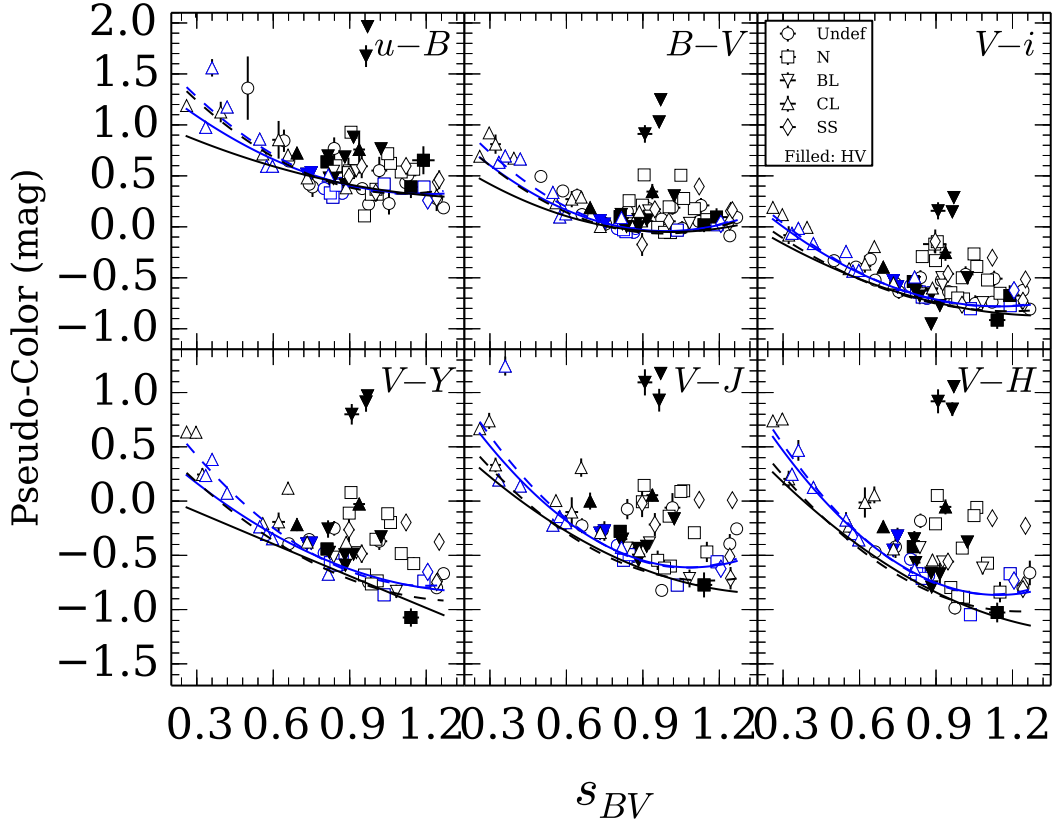}
\caption{Observed rest-frame colors, corrected
for foreground MW extinction, are plotted versus color-stretch. The meaning of the
symbols is the same as in Figure \ref{fig:BVmaxdm15}. Several best-fit
models for the intrinsic color loci are plotted as lines. Solid lines are
fits to the slow sample only, while dashed lines are fits to the entire sample.
Black lines are fits using the Cauchy prior, while blue lines are fits using the LRS
prior. \label{fig:color_sBV}}
\end{figure*}

\begin{figure*}
\includegraphics[width=6in]{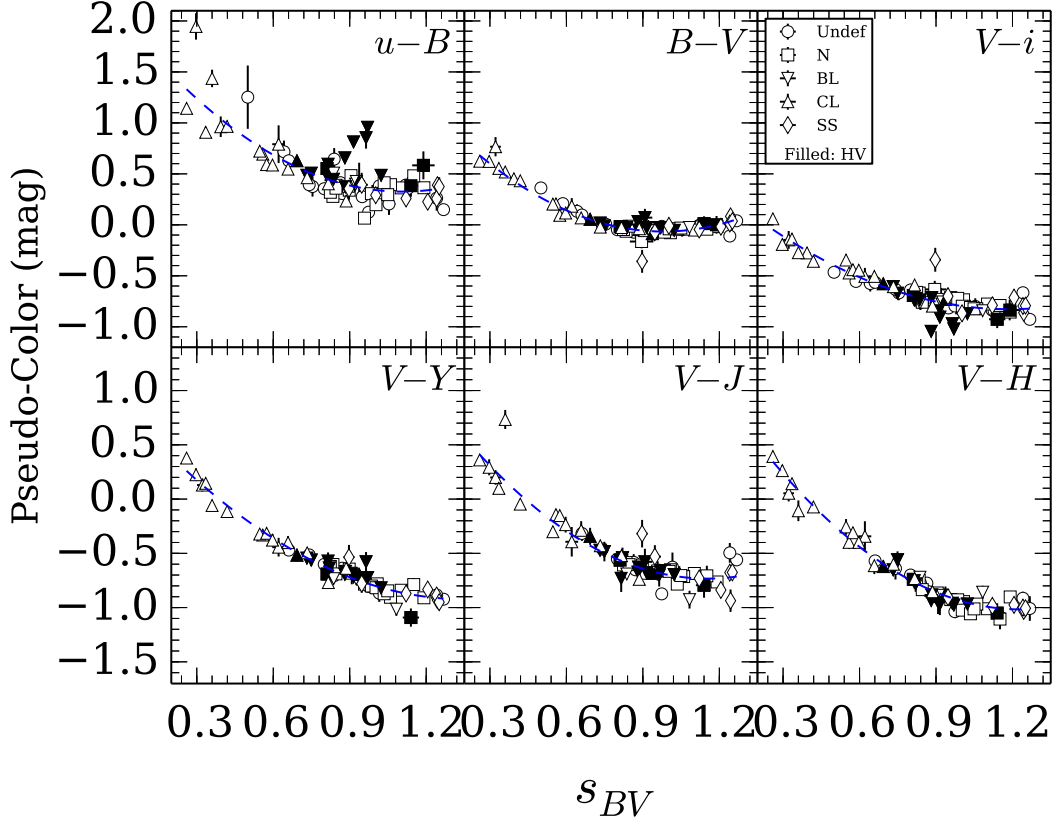}
\caption{Same as Figure \ref{fig:color_sBV}, but with both foreground MW and host galaxy 
extinction removed. The extinctions derived from the model with the Cauchy prior were
used and the best-fit intrinsic color loci are plotted as a dashed blue line.
\label{fig:color_sBV_corr}}
\end{figure*}

\subsection{Results}

In general, the Bayesian method we have developed does a good job
of finding the locus of bluest colors in our sample. Figures
\ref{fig:color_sBV} and \ref{fig:color_sBV_corr} show the observed
and host galaxy extinction-corrected pseudo-colors, respectively.
Aside from investigating
the effects of using different priors on our model, we also investigated
the effects of including different subsamples of our data by filtering
on two observables: the color-stretch and $B-V$ pseudo-color. We produced 2 subsamples:
1) the ``slow'' sample consisting of all objects with $s_{BV}>0.5$,
which primarily excludes the SN~1991bg-like objects, and 2) the ``blue''
sample that excludes objects with $\left(B_{max}-V_{max}\right)>0.5$,
eliminating the most heavily reddened objects. We now proceed to describe
the results of fitting with different priors and samples.

\subsubsection{Order of the Color-Stretch Relation}

The first thing we determined was whether or not the intrinsic color-stretch
relations, $P_{j}^{N}(s_{BV})$, require a linear or quadratic relation.
To accomplish this, we introduce a selection function $S(p_{1},p_{2})$
into our likelihood  which selects a linear color relation,
($N=1$) with probability $p_{1}$ and the quadratic color relation
($N=2$), with probability $p_{2}$. In this way, the MCMC chains
will spend some time trying the linear model and some time trying
the quadratic. We can compute the posterior probability $P(N)$ by
counting nodes in the chain. From that we can compute the odds ratio:
\begin{equation}
O=\frac{P(N=2)}{P(N=1)}\frac{p_{1}}{p_{2}}.\label{eq:BayesFactor}
\end{equation}
The odds ratio gives a measure of the likelihood of one model versus
the other, giving us the ability to select one if the data so warrant.
Note that the probabilities $P$ in Equation (\ref{eq:BayesFactor}) are computed
by marginalizing over all parameters and so there is a built-in ``Occam
factor'' that penalizes the $N=2$ model as it has one more free
parameter \citep{Gregory:2011}. Several pilot runs of our MCMC
gave odds factors between 16 and 88 in favor of the quadratic model.
The lower values of the odds ratio occur when we do not use the LRS
prior and restrict the data to $s_{BV}>0.5$. This can easily be seen
in Figure \ref{fig:color_sBV} where fits to the data without the
``low reddening'' prior tend to have less curvature than when the
``low reddening'' sample are constrained to have zero reddening.
In either case, an odds ratio of greater than 10 is considered reasonably
strong evidence \citep{Jeffreys:1961} to favor the quadratic model
over the linear one and we continue throughout with quadratic solutions
for the SN~Ia intrinsic colors.

\begin{figure}
\includegraphics[width=3.5in]{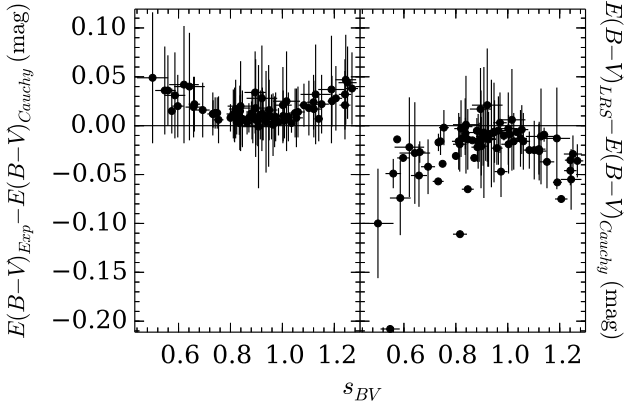}
\caption{Comparison of derived color excesses $E(B-V)_{host}$. In the left
panel, we plot the difference between $E(B-V)_{host}$ inferred using 
the exponential and Cauchy priors versus the color-stretch $s_{BV}$. 
In the right panel, we compare the Cauchy and LRS
priors. The error bars are the 68\% confidence intervals determined
from the posterior distributions for each quantity. \label{fig:EBV_priors}}
\end{figure}

\subsubsection{The $E(B-V)_{host}$ Prior}

When solving for the intrinsic colors, the factor that has the largest
effect is whether or not we use the LRS prior.
In most cases, $P_{j}^{N}$ ends up having a higher curvature when
we use the low reddening sample. This is most notable in the optical
minus NIR colors
$V-Y$, $V-J$, and $V-H$, where there are a significant number of
objects at high color-stretch that are bluer than the low reddening sample
and therefore tend to straighten out the intrinsic color-curves. This
is easily understood as the LRS prior, being associated with specific
objects, implicitly defines a color-stretch dependence on the intrinsic
colors, whereas the Cauchy and exponential priors apply to the sample
as a whole and are ``color-stretch blind''. At the high-
and low-color-stretch ends of the distribution, there are relatively few
objects and so the solution for the Cauchy and exponential prior has more freedom
at those ends. 

The resulting $E(B-V)_{host}$ color excesses also differ the most when comparing
the LRS prior to the analytic priors. The left panel of 
Figure \ref{fig:EBV_priors}
shows the difference between the color excesses from the exponential and Cauchy
priors. There is a systematic offset 
of about 0.02 mag in the sense that the exponential prior produces higher
extinctions and there is a systematic difference with $s_{BV}$, though both
effects are less than the 1-$\sigma$ errors in the extinctions themselves.
The right panel of Figure \ref{fig:EBV_priors} 
shows the difference between the color excesses
from the Cauchy and LRS priors. One can see there is a much larger
offset in the sense that the LRS prior produces lower
color excesses and there is a more pronounced systematic with respect
to $s_{BV}$. This is also easily understood as Figure \ref{fig:color_sBV}
shows that the intrinsic colors from the LRS are significantly redder
than those from the Cauchy or exponential priors. In both cases, the
systematic trend in the differences with $s_{BV}$
is due to low- and high-$s_{BV}$ ends having fewer objects so that the
choice of prior has more of an effect.

Along with varying the priors, we also varied the sample used to fit
the intrinsic colors. In particular, we were interested in whether
the fast-declining objects could be fit together with the normal Ia's.
Figure \ref{fig:color_sBV} shows the best-fit intrinsic
colors obtained when excluding (solid line) and
including (dashed line) the fast-declining SNe~Ia. The fits
are remarkably close. Indeed, when using the color-stretch parameter $s_{BV}$ 
instead of $\Delta m_{15}$, the fast-declining objects seem to lie on
a smooth extension to the color loci of the normal objects. 

For reasons stated above, we prefer to be agnostic about the reddening
and go forward with the results obtained without using the LRS. We
also adopt the Cauchy prior for the reddening as it allows a
higher probability for larger reddenings.
Table \ref{tab:color_coef}
lists the fits to the intrinsic color coefficients when using
different subsamples and assuming different priors. It also
tabuates the intrinsic scatter in the colors. These vary between
0.05 and 0.15 mag. The optical colors $B-V$ and $g-r$ have the
lowest intrinsic widths, followed by the optical-NIR colors. The
$u-B$ color stands out as having a partcularly large scatter.
We consider the results for the reddening
law in the next sections.

\subsubsection{Fitting a Single Reddening Law}
\label{sec:SingleRv}

If we use a simple uniform prior for $R_{V}$ that
is universal for the entire sample of SNe~Ia, we find that the recovered
value depends on the subsample of objects we fit and the form of the
reddening law. Table \ref{tab:Rv_global} lists the resulting values of
$R_V$ for these different cases.
The sensitivity to subsample was seen in our earlier work (F10) and
continues with our larger sample of objects. Generally speaking, when
we include those SNe~Ia with pseudo-colors redder than $B_{max}-V_{max}\simeq0.5$,
the resulting value for $R_{V}$ tends to be lower.

In all cases, the
F99 reddening law results in higher $R_V$ than CCM+O.
This is simply due to the different behaviors of either reddening
law as we extrapolate to $R_V < 2.6$.
All we can say is that when the reddest
SNe~Ia are removed from the analysis, the resulting value of $R_{V}=2.15\pm0.16$
for the F99 reddening law
is well below the typical value for the MW. For the remainder, we will
primarily use F99 to describe ISM dust, however we also include
results using CCM+O in our tables for comparison and in order to
be consistent with previous publications.

The dependence of $R_V$ on subsample was seen in F10 and it was argued 
that a possible explanation
was that the reddest SNe~Ia could have increased extinction due to
the proximity of dust local to the progenitor system (G08).
Indeed, a detailed analysis in F10 of the extinction as a function
of wavelength seemed to favor the G08 power-law over the standard
CCM+O law due to its ability to better fit the $u$-band extinction.
We therefore also attempt to fit the G08 reddening
law:
\begin{equation}
\frac{A_{j}}{A_{V}}=1-a+a\left(\frac{\lambda_{j}}{\lambda_{V}}\right)^{-p},
\end{equation}
where $\lambda_{j}$ is the effective wavelength of filter $j$. The
differential extinction is therefore
\begin{equation}
\Delta A_{j}=A_{V}a\left(\left(\frac{\lambda_{B}}{\lambda_{V}}\right)^{-p}-\left(\frac{\lambda_{j}}{\lambda_{V}}\right)^{-p}\right),
\end{equation}
where we see that $A_{V}$ and $a$ are completely degenerate. We
can therefore only make inferences on the product of these two parameters.
Column 4 of Table \ref{tab:Rv_global} lists the best-fit values for
the power index $p$.

\begin{figure}
\includegraphics[width=3.5in]{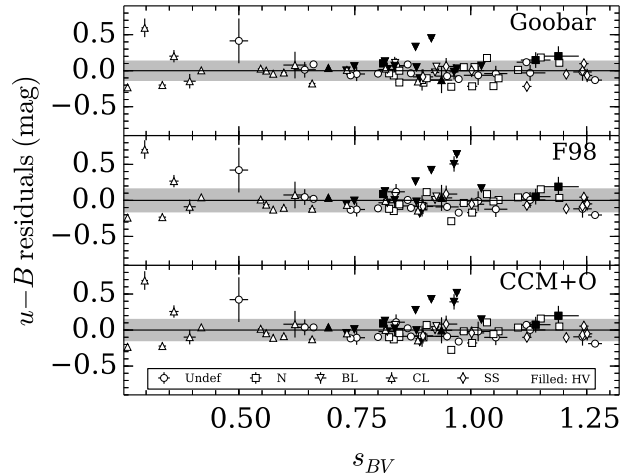}
\caption{The residuals in $u-B$ for the three 
different reddening laws
used in this paper. The gray shaded regions denote the intrinsic dispersion
in the $u-B$ color. The symbols have the same meaning as Figure \ref{fig:BVmaxdm15}.
Four objects with particularly high residuals are labeled.
\label{fig:u_resids}}
\end{figure}

As with fitting for $R_{V}$, the resulting values of $p$ depend
quite a bit on the data. We fit with the same 4 scenarios as before.
As with CCM+O and F99, removing the reddest objects results in a less steep 
reddening law,
and there is a slight dependence on inclusion of the
$u$ band.

\begin{figure}
\includegraphics[width=3.5in]{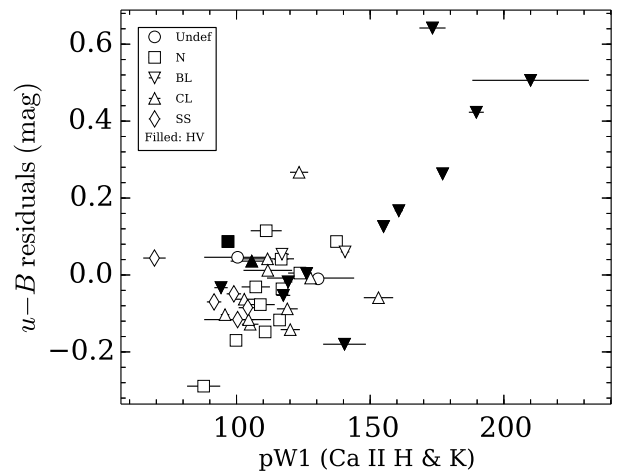}
\caption{The residuals in $u-B$ for the 
F99 reddening law versus the pseudo-equivalent width of \ion{Ca}{2} H \& K lines
from \citet{Folatelli:2013}. The BL HV objects with the highest $pW1$ are labeled.
The symbols have the same meaning as Figure \ref{fig:BVmaxdm15}.
Several BL-HV objects are labeled.
\label{fig:u_resids_pw1}}
\end{figure}

While it remains true that the $u$-band color excess for the two
reddest objects $ $(SN~2005A and SN~2006X) are better fit by the G08
power-law, these seem to be the only two and it may simply be a statistical
fluke. Figure \ref{fig:u_resids} shows the residuals in the color
model for $u-B$. Clearly, SN~2005A and SN~2006X show a better fit,
but very few others show a similar improvement. SN~2006X is known to have CSM
\citep{Wang:2008},
which was the motivation for the G08 power-law, so it
is interesting that it fits better. However, SN~2007le is also known
to have CSM \citep{Simon:2009}, and yet it is fit
equally well by CCM+O and F99. It is also conspicuous that the largest
outliers are all spectroscopically classified as broad line (BL) and
high-velocity (HV). This could indicate that it is spectral features
that are responsible for the non-standard $u-B$ colors. To investigate this,
we plot the $u-B$ residuals as a function of the pseudo-equivalent 
width of the \ion{Ca}{2} H \& K lines, $pW1$ from \citet{Folatelli:2013}.
This is shown in Figure \ref{fig:u_resids_pw1}. Clearly, the objects with
high $pW1$ have the highest residuals. We therefore conclude that
the anomalous $u-B$ colors of these BL objects are due to their
prominent \ion{Ca}{2} lines rather than an anomalous reddening law
due to CSM.

As a check, we can easily estimate the change in $u-B$ color due to a change
in $pW1$. We start by measuring synthetic $u$ and $B$ flux from the
\citet{Hsiao:2007} SED. We then artificially remove the \ion{Ca}{2} H \& K
lines by interpolating the continuum. The flux is then re-measured and a
flux decrement in each filter is computed. This flux decrement is scaled
from the \citet{Hsiao:2007} $pW1 = 108$\AA\ to $pW1 = 200$\AA\ and removed from
the original $u$ and $B$ fluxes.  The resulting change in $u$ is 0.33 mag fainter,
while the change in $B$ is 0.03 mag fainter, so that the change in $u-B$ is
0.3 mag redder, consistent with the effect seen.

In a recent paper, \citet{Mandel:2014} presented
evidence that BL-HV events show intrinsic color differences in $B-V$ and $B-R$ of
$\sim0.05-0.10$ mag with respect to normal SNe~Ia, and that these differences
originate in the $B$ band.  The intrinsic colors for the CSP sample plotted in
Figure \ref{fig:color_sBV} show similar deviations in $B-V$ for the same BL-HV events in our
sample that display anomalous u-B colors.   \citet{Foley:2011b} have argued
that this is a natural expectation for HV objects since for saturated lines, as
ejecta velocity increases, the width of the line will also increase.  This, in
turn, leads to larger equivalent widths and generally higher opacity in
wavelength regions where there are many strong lines.  This effect is
particularly strong in $u$ due to the strength of the \ion{Ca}{2} H \& K absorption.

This trend with \ion{Ca}{2} can also be seen in the work by \citet{Chotard:2011},
who modeled a reddening law based on spectrosopic obserations of SNe~Ia. After
correcting for decline rate, the reddening law had sharp features due to variaions
in the strength of
\ion{Ca}{2} H \& K, \ion{Ca}{2} NIR triplet, and \ion{Si}{2}~$\lambda 6355$. They apply
a color correction based on the \ion{Ca}{2} H \& K strength, which is precisely what
we see in Figure \ref{fig:u_resids_pw1}.

\begin{figure*}
\includegraphics[width=6in]{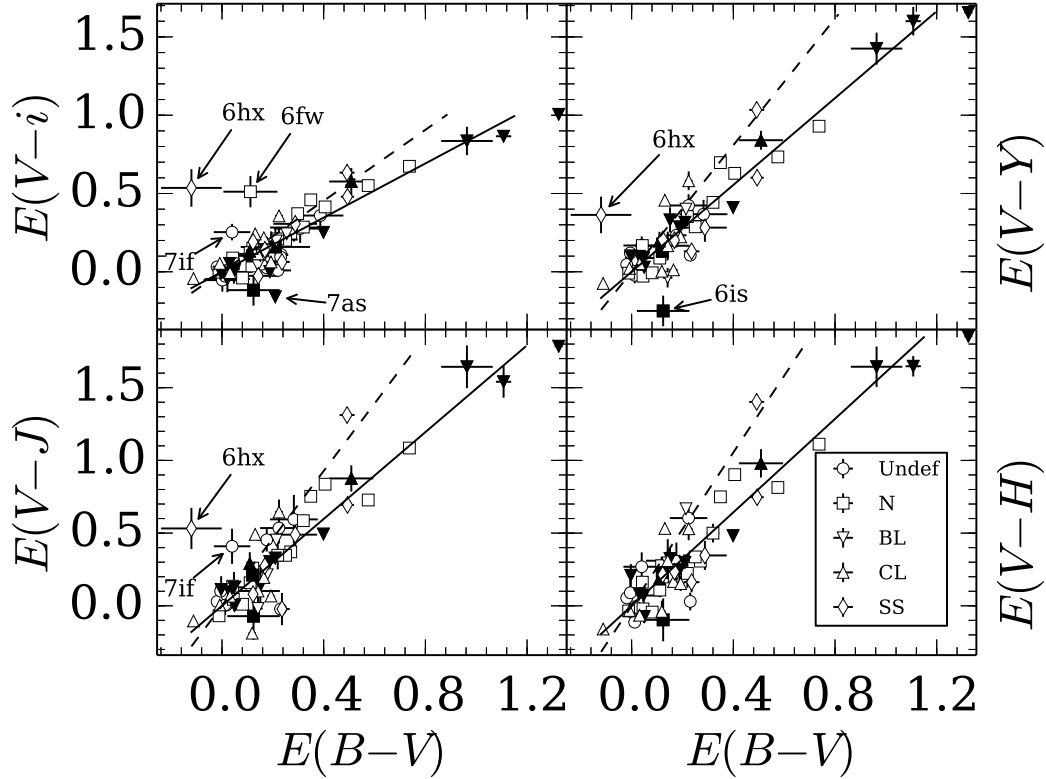}
\caption{Montage of several color excess vs. color excess plots. Each SN~Ia is a single
point in these diagrams. The meaning of the
symbols is the same as in Figure \ref{fig:BVmaxdm15}.  The $V$-NIR color 
extinctions are labeled on the
y-axes. Two representative reddening laws are plotted in each panel:
$R_{V}=3.1$ is plotted as a dashed line, while $R_{V}=1.7$
is plotted as a solid line. Objects with peculiar color excesses are labeled.
\label{fig:extinctions}}
\end{figure*}

\subsubsection{Fitting Individual $R_{V}$}

Given that the derived value of $R_{V}$ depends on the sample used,
it is natural to ask whether certain objects are driving
the solution, particularly those with larger color excesses which have
a greater ``pull''. In Figure \ref{fig:extinctions} we show a montage
of the extinctions in different NIR filters relative to $E(B-V)_{host}$.
In these extinction-extinction plots, a fixed reddening law, $R_{V}$,
should produce a correlation with fixed slope equal to $(R_{X}-R_{Y})$
where $X$ and $Y$ correspond to the particular filter combination
and the $R_{\lambda}$ values are dependent on the value of $R_{V}$
through the reddening law. Clearly, there is a large spread in the
correlation with a tendency to small values of $R_{V.}$ Because of this,
we allow the reddening coefficient $R_{V}$ to vary for each SN~Ia
in the sample.

As a first step, we again use uniform priors on the individual $R_{V}$.
For objects with very little extinction or no NIR photometry, $R_{V}$
will be very poorly constrained. Nevertheless, the inferred $E(B-V)_{host}$
and intrinsic color loci \emph{will} be well-defined. We are primarily
interested in those objects for which $R_{V}$ can be constrained
and require only uniform priors.
 
\begin{figure}
\plotone{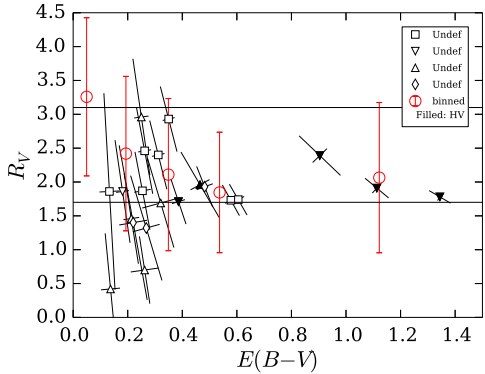}
\caption{The best-fit $R_V$ as a function of the best-fit color excess $E(B-V)_{host}$
for each individual SN~Ia using the Cauchy prior. Only objects for
which $R_{V}$ was significantly constrained are plotted. The error-bars
are drawn to show the correlation between the two variables and correspond
to the principal axes of the 1-$\sigma$ error ellipse for each point.
The meaning of the symbols is the same as Figure \ref{fig:BVmaxdm15}.
The two horizontal lines show the typical value $R_{V}=1.7$ derived
for SNe~Ia and the canonical value $R_{V}=3.1$ for the Milky-Way.
The mean and standard deviations of the binned prior from 
\S \ref{sec:Rv_priors} are plotted as red points and error-bars.
\label{fig:EBV_Rv}}
\end{figure}

\begin{figure*}
\plotone{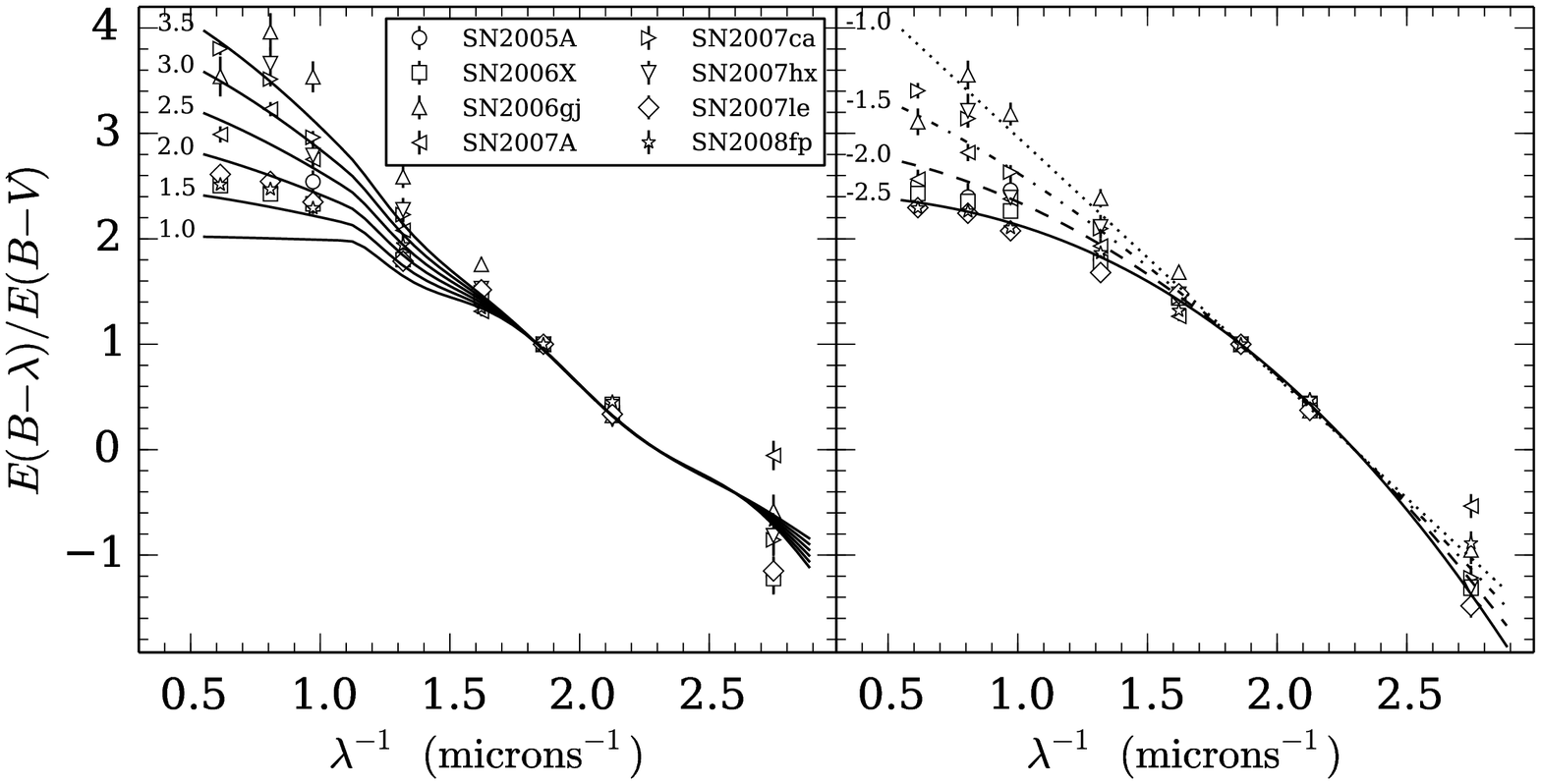}
\caption{The color excesses $E(B-\lambda)$ normalized
by $E(B-V)_{host}$ as a function of inverse wavelength for a sample of 8 SNe~Ia with
a range of reddening laws. In the left panel, the solid lines
are CCM+O laws with different values of $R_{V}$ as labeled to the left
of each curve. In the right panel, G08 power-law fits are plotted
instead. The value of $p$ is labeled to the left of each line. \label{fig:EB-lamb}}
\end{figure*}

The results of fitting $E(B-V)_{host}$ and
$R_V$ are shown in Figure \ref{fig:EBV_Rv} and tabulated in
columns 6, 7, and 10 of Table \ref{tab:SNe}. As expected, the
value of $R_{V}$ becomes unconstrained as the amount of reddening
tends to zero. In general, it would seem that the objects with moderate
color excesses ($0.2<E(B-V)_{host}<0.5$) have a range of $R_{V}$ from the rather
low values that have been endemic of SN~Ia studies to values more
in line with the canonical Milky-Way value of $R_{V}=3.1$, though these
seem to be in the minority.  
All the more reddened objects tend to the low value of $R_{V}\sim1.7$.
Figure \ref{fig:EB-lamb} shows a plot similar to Figure 14 in F10
where we plot the color excess in different filters as a function
of inverse wavelength. We have normalized the color excesses by $E(B-V)_{host}$
in order to focus on the shape of the extinction curve. CCM+O curves
with a range of values of $R_{V}$ are shown for reference. Clearly,
the colors of some objects like SN~2006gj are consistent with
values closer to the Milky-Way
value of $R_{V}=3.1$ whereas colors of objects like SN~2007le would imply a lower
value of $R_{V}\sim1.7$. Others have attempted to model a relationship
between $R_{V}$ and the observed colors \citep{Mandel:2011}, however
in the absence of any theoretical model to motivate such a relationship,
and significantly fewer objects in our sample,
we will not do so at this time. It is fair to say, however, that a
simple one-parameter description of the extinction suffered by SNe
Ia is insufficient to explain the diversity of inferred values of
$R_{V}$.

There are 4 objects which prefer very low values of $R_V$, 
such that they give us upper limits. These objects, SN~2004eo, SN~2005ke,
SN~2007ba, and SN~2008gp, have moderately red optical colors, but very blue
optical-NIR colors. These cannot be explained with any reasonable 
reddening law and must be due to intrinsically peculiar colors
or systematic errors in the NIR photometry.
We have already argued that SN~2005ke is a transitional
object between fast and normal SNe~Ia and SN~2007ba has a similar 
$s_{BV}$. They, and SN~2004eo, are classified as spectroscopic
type CL, while SN~2008gp is unclassified. However, there are other
CL types that have ranges of $R_V$ that are perfectly normal. 
Unfortunately, we do not possess NIR spectra of these objects and
cannot determine whether there are spectroscopic peculiarities in the
NIR.

\subsubsection{Inferring Milky-Way Extinctions}

Throughout this paper, we have assumed foreground MW color excesses
from \citet{Schlafly:2011}. While we include the error for these
measurements in our analysis, one might be concerned that there may
be a systematic difference between the true color excess and
the \citet{Schlafly:2011} values. In order to test this, we can look
at objects whose host galaxy extinctions are expected to be low
(due to host type, projected distance, and lack of \ion{Na}{1}
absorption), yet have significant MW extinction according to
\citet{Schlafly:2011}.

There are three such objects in our sample:  SN~2006kf ($E(B-V)_{MW} = 0.210$),
SN~2008bc ($E(B-V)_{MW} = 0.225$), and SN~2008ia ($E(B-V)_{MW} = 0.195$).
In a trial run, these objects' foreground extinctions were artificially
set to zero, allowing us to determine the foreground extinction based
on the observed colors of the SNe rather than extinction maps. The 
results are given in Table \ref{tab:MW_EBV}. The values of $E(B-V)_{host}$
are quite close and show no sign of significant systematic 
difference, though it is difficult to be certain with only three
objects. The derived values of $R_V$ are consistent with the
spread in values associated with MW sight-lines \citep{Cardelli:1989,Fitzpatrick:1999}.

\subsubsection{General $R_{V}$ Priors}
\label{sec:Rv_priors}
We finish with an attempt to construct a prior for $R_{V}$ that can be
used when either 1) the extinction is too low or 2) there is insufficient
filter coverage. We do this in two ways: by constructing a Gaussian mixture
model as done in \citet{Kelly:2007} for all objects and splitting the objects
into bins of $E(B-V)_{host}$ and constructing an independent Gaussian distribution
for each bin. 

The Gaussian mixture model is simply a sum
of $N$ Gaussians, forming a composite prior that can have larger
wings or multiple peaks. Details of the construction of this prior
can be found in the Appendix and \citet{Kelly:2007}. 
We run our MCMC analysis using the same sub-samples and filter sets
as in \S \ref{sec:SingleRv}. We find that two Gaussian
components are sufficient, though the second is barely significant.
Table \ref{tab:Ngauss} summarizes
the values of the mixture model's hyper-parameters. It is evident
that the parameters are quite insensitive to which subsample and
filter set we use.  We can therefore use any to represent a prior
on $R_V$ for future studies.

Our second approach is to bin the data by $E(B-V)_{host}$
and solve for an independent single Gaussian prior for each bin. The
hope is that the larger numbers of objects at low color excesses will
balance their relatively weak pull on the data and that the reddest
objects will not overly influence their prior. We split the data into
5 bins. The results are given in Table \ref{tab:Rv_binned} and plotted
in Figure \ref{fig:EBV_Rv} as red circles denoting the mean and
error bars denoting the standard deviation of each prior. Clearly,
the average $R_{V}$ diminishes with $E(B-V)_{host}$, starting with typical
$R_{V}\sim3-4$ found in the MW for the least reddened objects. 
It is hard to imagine a scenario in which such a correlation should
exist and could simply be due to low numbers highly-reddened objects
and biases in follow-up selection (e.g., highly reddened, fainter objects
close to their host galaxy are selected against for
spectroscopic follow-up and typing). One should also note that all the 
BL and HV objects have low
$R_V$, while the other objects show more of a spread. Deciding whether this is
due to differences in ISM environment or spectroscopic diversity (SN physics)
will require more objects with moderate to high color excesses in a variety
of locations within their hosts.

The values of $R_V$ for each object that results from the mixture model prior
and binned prior
are tabulated in columns 8, 9, 11, and 12 of Table \ref{tab:SNe}.

\section{Summary}
\label{sec:summary}

This latest analysis paper by the CSP has focused primarily on the
intrinsic colors of SNe~Ia and what they tell us about the population.
In particular, we have shown that when attempting to categorize the
SN, the choice of light-curve parameter is an important one when
it comes to the fastest-evolving objects. We find that the traditional
observable $\Delta m_{15}(B)$ is not a very reliable parameter when
$\Delta m_{15}(B)>1.7$. At this point, several photometric characteristics 
become
uncorrelated with $\Delta m_{15}(B)$ and it therefore loses its power
to predict the light-curve behavior as well as the intrinsic colors
of the objects. 

We offer a new definition of an old parameter, namely the color-stretch
$s$. In order to apply it to the faster-evolving objects, whose light-curves
are sufficiently different from ``normal'' SNe~Ia, we define this
``new'' color-stretch $s_{BV}$ to be the time of rest-frame $B-V$ maximum
relative to $B$ maximum divided by 30 days. This produces a stretch
parameter that is sufficiently close to the parameter used by other
groups yet being solely based on observed light-curve behavior, can
be measured accurately for fast-evolving SNe~Ia that have sufficient
coverage in $B$ and $V$. When using this light-curve parameter rather than
$\Delta m_{15}(B)$, the fastest evolving SNe~Ia appear less as a distinct
population of objects and more as the tail end of normal SNe~Ia.

With an increased number of objects, we have solved for the intrinsic
colors of SNe~Ia as a function of $s_{BV}$. We present two approaches
to determining the blue locus: 1) singling out a subsample that are
believed to be have low reddening, and 2) assigning a prior to the reddening
that is peaked at zero and has a long tail to higher color excesses. In
both cases, we find that a quadratic function of color-stretch is required
to fit the intrinsic colors. We also find that the  LRS prior
favors color functions with more curvature, particularly in the NIR
and redder intrinsic colors. This is due to a significant number of
high-$s_{BV}$ objects having significantly bluer colors than the LRS objects.
We therefore choose not to use the LRS to
determine the intrinsic colors. 

We continue to find that the value of $R_{V}$ favored by SNe~Ia tends
to be lower than the usual value found in the Milky-Way and other
external galaxies, though with a large range in values 
\citep[see e.g.,][]{Krisciunas:2000,Astier:2006,EliasRosa:2006,EliasRosa:2008,Wood-Vasey:2008,Mandel:2011}. This would
appear to be in contrast
to results from \citet{Chotard:2011}, who arrive at a 
``normal'' value of $R_V$
when analysing the spectra of SNe~Ia rather than photometry,
and to the results of \citet{Scolnic:2014}, who find that
one biases $R_V$ to artificially low values
when intrinsic color variations are not
taken into account.  However in both cases, the objects 
considered have predominantly lower reddenings than our
sample, corresponding to $E(B-V)_{host} < 0.2$. This corresponds
to the first two bins in our binned $R_V$ prior, for which we
obtain an average $R_V \simeq 3.1$. 
Further, spectropolarimetric 
work on SNe~Ia indicate
that the interstellar dust responsible for the extinction  has
a range of $R_V$ values \citep{Hough:1987,Wang:2003,Patat:2009},
including values as low as those derived here.
Unless there are two different systematics at work
that are conspiring to lower $R_V$, we must conclude
that these abnormal reddenings are real.

The
fact that some SNe~Ia seem to have a different reddening law than
typical interstellar gas could indicate that those objets are obscured
by a local environment with different properties than is typical for
the MW ISM. It is interesting to note that objects with highest $E(B-V)_{host}$
and lowest $R_{V}$ are predominantly high-velocity objects, which
have been shown to reside preferentially in the haloes of galaxies
\citep{Wang:2013}. This fits in well with observations of stars at
high Galactic latitudes that are better fit by a low value of
$R_V$ \citep{Szomoru:1999}.

It has also been suggested that the abnormally low values of $R_{V}$
could be due to CSM \citep{Wang2005,Goobar:2008} and
while two of our objects (SN~2006X and SN~2005A) have $u$-band colors
that are more consistent with a G08 power-law model for CSM
several other objects, including SN~2007le which is known
to have CSM, are also consistent with the CCM+O and F99
reddening laws. The spectropolarimetric data would also support
ISM over CSM.  Finally,
high-resolution spectral observations of SNe~Ia show that the equivalent
width of diffuse interstellar bands at the systemic redshift of the
host is well correlated with the reddening inferred from photometry
\citep{Phillips:2013}, implying that reddening is not related to dust
in the immediate environment of SNe~Ia.

We have shown that there is an intrinsic spread in the shape of the
reddening laws for SNe~Ia. The typical practice of fitting for a
single $R_{V}$ for all SNe~Ia therefore underestimates the error
incurred by an intrinsic parent distribution with finite width. We have
constructed a model for this distribution that could
be used in cosmological fitting and distance determinations when full
optical {\it and}  NIR photometric coverage is not available and/or when the
reddening is low. The simplest, being a Gaussian mixture model,
captures the intrinsic spread in $R_V$ and can easily be used
as a prior in cosmological and distance estimations. The prior
binned in $E(B-V)_{host}$ also provides a useful prior and is perhaps more
appropriate for use when considering individual objects.

The NIR continues to offer an attractive way to circumvent the issues
of extinction. As would be expected for obscuration by dust, the longer
wavelengths have much lower color corrections. And while the Phillips
relations in the NIR are non-zero \citep{Krisciunas:2004,Wood-Vasey:2008,Kattner:2012},
they are significantly lower
than in the optical wavelengths.
This has encouraged the CSP to further
investigate SNe~Ia in the NIR and we are mid-way through
a 4-year campaign
at Las Campanas Observatory to obtain optical and NIR photometry of approximately
100 SNe~Ia in the Hubble flow ($0.01<z<0.1$). This will reduce the
contribution of peculiar velocities to the scatter in the NIR Hubble
diagram, allowing us to measure the intrinsic scatter in their
luminosities more
reliably. An increase in the nearer sample size, particularly objects
that suffer from larger extinctions, will also help to constrain the
distribution of dust
properties.

\acknowledgements{}

The work of the CSP has been supported by the National Science Foundation under
grants AST0306969, AST0607438, and AST1008343. M.~D. Stritzinger and C.
Contreras gratefully acknowledge generous support provided by the Danish Agency
for Science and Technology and Innovation  realized through a Sapere Aude Level
2 grant. Computing resources used for this work were made possible by a grant
from the Ahmanson Foundation. This research has made use of the NASA/IPAC
Extragalactic Database (NED) which is operated by the Jet Propulsion
Laboratory, California Institute of Technology, under contract with the
National Aeronautics and Space Administration.
\clearpage

\appendix{}

\section{Markov Chain Monte Carlo}\label{sec:Appendix1}

In this appendix we detail the model used to solve for the intrinsic
colors, color excesses, and reddening laws for our sample. In the Bayesian
framework, one computes the probability that the data is observed,
given the parameters of the model. We denote this as $p\left(D\left|\theta\right.\right)$
where $D$ represents the data and $\theta$ represents the parameters.
We then also assign prior probabilities on the parameters themselves,
$p\left(\theta\right)$. These priors are chosen to be appropriate
probability distributions and can themselves be functions of parameters
(commonly referred to as hyper-parameters). Bayes' theorem then states
that the posterior probability of the parameters is given by:
\begin{equation}
p\left(\theta\left|D\right.\right)=\frac{p\left(D\left|\theta\right.\right)p\left(\theta\right)}{\int d\theta p\left(D\left|\theta\right.\right)p\left(\theta\right)}
\end{equation}
Inference is done by finding the mode and moments of this equation,
which must be done numerically for all but the simplest models. A
very popular method today is Markov Chain Monte Carlo (MCMC). MCMC
works by creating a Markov Chain of parameter states $\{\theta_{i}\}$
in which $\theta_{i}$ only depends on the previous state $\theta_{i-1}$.
The sampling method dictates how one goes from state $\theta_{i-1}$
to $\theta_{i}$ and there are several popular alternatives. We use
the Metropolis-Hastings method in which one starts at an initial state
$\theta_{i}$ and a new state $\theta^{\prime}$ is proposed:
\begin{equation}
\theta^{\prime}=\theta_{i}+\delta\theta,\ \ \delta\theta\sim N(\theta_{i},C_{MH})
\end{equation}
where the perturbing vector $\delta\theta$ is drawn from a multivariate
normal distribution with covariance matrix $C_{MH}$. The proposed
state is accepted with probability
\begin{equation}
p\left(\theta^{\prime}\left|\theta_{i}\right.\right)=\min\left(1,\frac{p\left(\theta^{\prime}\right)}{p\left(\theta_{i}\right)}\right)\label{eq:Metropolis-Hastings}
\end{equation}
in which case $\theta_{i+1}=\theta^{\prime}$. If the proposed state
is rejected, $\theta_{i+1}$ is assigned a copy of $\theta_{i}$.
In this way, the Markov chain always migrates to a region of higher
probability; however once there, it explores the region in a random
fashion, though constrained by the shape of the probability distribution.
After a sufficient burn-in period, sampling from the Markov chain
is equivalent to sampling from $p\left(\theta\left|D\right.\right)$.
Inference on the parameters of interest can therefore be done by performing
statistics on the Markov Chain.

One crucial aspect of the Metropolis-Hastings method is the choice
of covariance matrix $C_{MH}$ used to propose the next state in the
Markov Chain. As one can see from Equation (\ref{eq:Metropolis-Hastings}),
if one chooses steps that are too large (or in directions away from
higher probability), the Markov chain can get ``stuck'' in place
for long periods of time. Conversely, if the steps are too small,
it can take a long time to converge to a region of high probability.
Choosing the appropriate $C_{MH}$ is therefore required to have an
efficient MCMC. We use an adaptive Metropolis-Hastings sampler that
begins with a simple diagonal $C_{MH}$ with variances chosen to reflect
the scales of the parameters. After enough proposals are accepted,
an empirical covariance matrix is computed from the chain up to that
point. In this way, the MCMC sampler becomes more efficient. The numerical
machinery for all this is built in to the Python package pymc \citep{Patil:2010},
which we use for all our modeling.

In order to ensure convergence, we run 4 parallel Markov chains for
each simulation and compute the Gelman-Rubin statistic $R$ \citep{GelmanRubin92}
for every parameter. This is essentially a measure of the ratio of
the mean of the variances of each chain to the variance of all chains
combined.

\subsection{Bayesian Hierarchical Model}

In this section, we outline the Bayesian Hierarchical model used to
infer the intrinsic colors, color excesses, and reddening laws for our
sample of supernovae. The model is termed hierarchical because there
is a hierarchy of parameters. We have the parameters we are most interested
in, and then we have the hyper-parameters that control the shapes
of the priors we impose on our parameters of interest. For example,
in one model we impose a Normal prior on $R_{V}$. This normal distribution
has a mean $\mu$ and variance $\sigma^{2}$ whose values we do not
know \emph{a priori}. All three are parameters, but $\mu$ and $\sigma^{2}$
are termed hyper-parameters. We could even choose to impose a prior
on $\sigma^{2}$ and have that prior depend on yet more hyper-parameters. 

In the case of our model, we use uniform priors wherever possible
and reserve more complicated priors for $E(B-V)_{host}$, $R_{V}$, and intrinsic
variances. We begin by defining the probability of our data, given
the parameters. We define the observable data as the set $D_{i}=\left\{ B_{i}-m_{i,j},s_{BV,i},E(B-V)_{MW,i}\right\} $
where for each SN (indexed by $i$), $B_{i}$ is the observed $B$-band
maximum and $m_{i,j}$ are the observed $ugriVYJH$ magnitudes at
maximum, $s_{BV,i}$ are the color-stretch parameters, and $E(B-V)_{MW,i}$
are the MW extinctions from \citet{Schlafly:2011}. We also define
$D_{i}^{\prime}$ as the ``true'' values of the observables. The
probability of the data is then 
\begin{equation}
\log p\left(D_{i}\left|D_{i}^{\prime}\right.\right)\sim\sum_{i}\left(D_{i}-D_{i}^{\prime}\right)^{T}C_{i}^{-1}\left(D_{i}-D_{i}^{\prime}\right)
\end{equation}
where $C_{i}$ are the covariance matrices that include errors in
the photometry, color-stretch, MW reddening and any associated correlations.
We also include an intrinsic variance $\sigma_{j}^{2}$ 
in $C_i$ for each distinct
color, which is left as a parameter to be determined. The ``true''
colors $B_{i}^{\prime}-m_{i,j}^{\prime}$ are given deterministically
by our model as a function of the parameters and ``true'' values
of the other observables:
\begin{equation}
B_{i}^{\prime}-m_{i,j}^{\prime}=P_{j}^{N}\left(s_{BV,i}^{\prime}-1\right)+\Delta A_{j}\left(E(B-V)_{host,i},R_{V,i}\right)+\Delta A_{j}\left(E(B-V)_{MW,i}^{\prime},3.1\right)
\end{equation}
while the other ``true'' values $s_{BV,i}^{\prime}$ and $E(B-V)_{MW,i}^{\prime}$
are free parameters. In this way, at each step of the Markov chain,
the sampler will perturb the values of the observables consistent
with the errors and covariances, thereby propagating their uncertainties
properly. The differential extinction $\Delta A_{j}=A_{B}-A_{j}$
for each filter $j$ and SN $i$ is a deterministic function of the
reddening coefficient $R_{V,i}$ and color excess $E(B-V)_{host,i}$. This
function is determined numerically by multiplying the \citet{Hsiao:2007}
SN~Ia SED at maximum with a CCM+O, F99, or G08 reddening law with the
specified $R_{V}$ and $E(B-V)_{host}$ and computing synthetic, reddened
photometry. Note that for the MW extinction, a constant $R_{V}=3.1$
is assumed.

\subsubsection{Priors}

With the probability of the data defined, all that remains is to specify
the priors. Wherever possible, we employ uniform priors, relying on
the data to constrain the parameters%
\footnote{In practice, the priors are uniform over a finite range of values
chosen to be much larger than the posterior probability distribution
for the parameter. %
}. This is done for the $N+1$ coefficients of the polynomial $P_{j}^{N}$
for each intrinsic color, the true values of the color-stretch $s_{BV,i}$,
and the MW color excess $E(B-V)_{MW,i}$. For the host galaxy
color excess, we use one of three possible priors. The first is a
conditional prior such that objects in the LRS
are assigned zero color excess and the rest are drawn from a uniform
prior:
\begin{eqnarray}
p\left(E(B-V)_{host,i}\right) & = & \delta(0),\ \ i\in\mathrm{LRS} \nonumber\\
 & = & U\left(0,\infty\right)i\notin\mathrm{LRS},
\end{eqnarray}
where $\delta()$ is the Dirac delta function.
The second is the same prior as used by \citet{Jha:2007}, namely
an exponential distribution with scale $\tau$:
\begin{equation}
p\left(E(B-V)_{host,i}\right)\propto\exp\left(-\frac{E(B-V)_{host,i}}{\tau}\right).
\end{equation}
Lastly, we investigate using a prior that is similar to the exponential
prior, but with longer tail, the truncated Cauchy distribution:
\begin{eqnarray}
p\left(E(B-V)_{host,i}|\tau,E_{max}\right) & = & \left[\tau\arctan\left(\frac{E_{max}}{\tau}\right)\left(1+\left(\frac{E(B-V)_{host,i}}{\tau}\right)^{2}\right)\right]^{-1},\ \ 0<E(B-V)_{host,i}<E(B-V)_{max} \nonumber\\
 & = & 0,\ \ \mathrm{otherwise},
\end{eqnarray}
where again $\tau$ represents a scale length and we impose a maximum
$E(B-V)_{max}$ to ensure finite probability. In both the exponential
and Cauchy priors, an inverse-gamma distribution is used for the hyper-parameter
$\tau$:
\begin{equation}
p\left(\tau\right)=\Gamma^{-1}\left(\alpha=1,\beta=0.2\right),
\end{equation}
where $\alpha$ and $\beta$ are chosen to give a relatively broad
prior peaked near the typical color excesses found in SNe~Ia.

\subsubsection{$R_{V}$ Priors}

In this paper we investigate several priors for $R_{V}$ that are
motivated by previous work done in the field. The first and simplest
is the assumption that there is a single universal $R_{V}$ for all
SNe~Ia. We only insist that it be strictly positive:
\begin{equation}
p\left(R_{V}\right)=U\left(0,\infty\right).
\end{equation}
We therefore solve for the most likely value for $R_{V}$ and its
error. This is the prior typically used in most light-curve fitters,
SNooPy included. A more realistic assumption would be that there is
some intrinsic distribution of $R_{V}$ and we instead try to infer
the properties of this distribution. Inspired by the work of \citet{Kelly:2007},
we use ``Gaussian mixture'' prior, which is a sum of Gaussians,
each with its own mean, standard deviation, and normalization:
\begin{equation}
p\left(R_{V,i}\left|\pi_{k},\mu_{k},\sigma_{k}^{2}\right.\right)\propto\sum_{k}\pi_{k}\exp\left(-\frac{1}{2\sigma_{k}^{2}}\left(R_{V,i}-\mu_{k}\right)^{2}\right).
\end{equation}
The hyper-parameters $\mu_{k}$, $\sigma_{k}^{2}$, and $\pi_{k}$
are themselves given the following hierarchical priors:
\begin{eqnarray}
p\left(\pi_{k}\right) & = & \mathrm{Dirichlet\left(1,\ldots,1\right),} \nonumber\\
p\left(\mu_{k}\left|\mu_{0},\sigma_{0}^{2}\right.\right) & = & N\left(\mu_{0},\sigma_{o}^{2}\right), \nonumber\\
p\left(\sigma_{k}^{2}\left|w^{2}\right.\right)=p\left(\sigma_{0}^{2}\left|w^{2}\right.\right) & = & \mathrm{Inv}\chi^{2}\left(1,w^{2}\right), \nonumber\\
p\left(\mu_{0}\right) & = & U\left(-\infty,\infty\right), \nonumber\\
p\left(w^{2}\right) & = & U\left(0,\infty\right),
\end{eqnarray}
where the Dirichlet prior simply ensures $\sum_{k}\pi_{k}=1$ and
$N$ is the normal distribution. Though complicated, this set of priors
simply ensures that the difference between the means of the Gaussians
$\mu_{k}$ are on the order of their widths $\sigma_{k}^{2}$. $ $The
parameter $\mu_{0}$ controls the overall mean of the composite distribution
while $w^{2}$ controls its overall width and separation of local
maxima.

The final prior we use reflects the findings of \citet{Mandel:2011},
namely that $R_{V}$ tends to be smaller on average for larger $E(B-V)_{host}$.
We therefore construct a prior in which we bin the data based on the
value of $E(B-V)_{host}$ and assign an independent prior for each bin. The
prior for each bin is taken to be a single Gaussian with independent
mean and standard deviation:
\begin{equation}
p\left(R_{V,i}\left|E(B-V)_{host,i},\mu_{k},\sigma_{k}^{2}\right.\right)=N\left(\mu_{k},\sigma_{k}^{2}\right),\; x_{k-1}<E(B-V)_{host,i}<x_{k},
\end{equation}
where the $x_{k}$ denote the bin boundaries.
\bibliographystyle{apj}
\bibliography{AnalysisII}

\LongTables
\begin{deluxetable*}{lclccccccccc}
\tablewidth{0pc}
\tablecolumns{12}
\tablecaption{SNIa Color Excesses and Reddening Laws\label{tab:SNe}}
\tablehead{
\colhead{}   & \colhead{Spec.}             & \colhead{}    & \colhead{}         & \colhead{}              & \colhead{}                
& \multicolumn{3}{c}{CCM+O} & \multicolumn{3}{c}{F99} \\
\cline{7-9} \cline{10-12} \\
\colhead{SN} & \colhead{Class.} & \colhead{HV?} & \colhead{$s_{BV}$} & \colhead{$E(B-V)_{MW}$} & \colhead{$E(B-V)_{HOST}$} 
& \colhead{$R_{V,U}$} & \colhead{$R_{V,B}$} & \colhead{$R_{V,M}$}
& \colhead{$R_{V,U}$} & \colhead{$R_{V,B}$} & \colhead{$R_{V,M}$}
}
\startdata
2004ef & BL & yes & 0.815(0.003) & 0.046(0.001)        & $0.158(0.024)$ & $2.7^{+1.5}_{-0.7}$ & $2.5^{+0.9}_{-0.6}$ & $1.8^{+0.7}_{-0.5}$ & $3.0^{+1.3}_{-0.6}$  & $2.7^{+0.8}_{-0.5}$  & $2.2^{+0.6}_{-0.5}$\\
2004eo & CL & no & 0.816(0.005) & 0.093(0.001)         & $0.128(0.024)$ & $<1.4$              & $0.6^{+1.2}_{-0.2}$ & $0.5^{+0.8}_{-0.2}$ & $0.8^{+0.8}_{-0.2}$  & $0.9^{+0.9}_{-0.2}$  & $0.8^{+0.7}_{-0.2}$\\
2004ey & N & no & 1.008(0.002) & 0.119(0.014)          & $0.019(0.020)$ & \ldots              & $3.1^{+2.3}_{-1.5}$ & $1.3^{+1.0}_{-0.5}$ & \ldots               & $3.1^{+1.7}_{-1.2}$  & $1.7^{+0.8}_{-0.6}$\\
2004gc & \ldots & no & 0.921(0.023) & 0.178(0.004)     & $0.242(0.052)$ & \ldots              & $2.0^{+1.1}_{-0.9}$ & $1.5^{+0.9}_{-0.6}$ & \ldots               & $2.3^{+1.1}_{-1.0}$  & $1.8^{+1.1}_{-0.6}$\\
2004gs & CL & yes & 0.693(0.004) & 0.026(0.001)        & $0.148(0.024)$ & $2.8^{+1.4}_{-0.7}$ & $2.5^{+0.8}_{-0.6}$ & $1.8^{+0.6}_{-0.4}$ & $3.0^{+1.4}_{-0.6}$  & $2.7^{+0.8}_{-0.5}$  & $2.3^{+0.6}_{-0.4}$\\
2004gu & SS & no & 1.244(0.010) & 0.022(0.001)         & $0.096(0.034)$ & \ldots              & $1.9^{+1.1}_{-0.8}$ & $1.4^{+0.7}_{-0.6}$ & \ldots               & $2.1^{+1.0}_{-0.7}$  & $1.8^{+0.7}_{-0.6}$\\
2005A & BL & yes & 0.964(0.010) & 0.026(0.001)         & $1.129(0.029)$ & $1.4^{+0.1}_{-0.1}$ & $1.4^{+0.1}_{-0.1}$ & $1.4^{+0.1}_{-0.1}$ & $1.9^{+0.1}_{-0.1}$  & $1.9^{+0.1}_{-0.1}$  & $1.8^{+0.1}_{-0.1}$\\
2005M & SS & no & 1.206(0.003) & 0.027(0.002)          & $0.060(0.021)$ & \ldots              & $3.2^{+2.1}_{-0.9}$ & $1.8^{+0.9}_{-0.5}$ & \ldots               & $3.4^{+1.7}_{-0.9}$  & $2.1^{+0.9}_{-0.5}$\\
2005W & BL & no & 0.923(0.009) & 0.061(0.001)          & $0.233(0.025)$ & \ldots              & $2.3^{+1.1}_{-0.9}$ & $1.6^{+0.9}_{-0.6}$ & \ldots               & $2.5^{+1.0}_{-0.8}$  & $2.0^{+0.7}_{-0.6}$\\
2005ag & BL & no & 1.083(0.008) & 0.033(0.001)         & $0.072(0.021)$ & \ldots              & $2.9^{+2.0}_{-0.9}$ & $1.6^{+0.8}_{-0.6}$ & \ldots               & $2.5^{+1.2}_{-0.7}$  & $2.0^{+0.8}_{-0.6}$\\
2005al & \ldots & no & 0.864(0.003) & 0.048(0.002)     & $0.022(0.013)$ & \ldots              & $4.0^{+2.5}_{-1.4}$ & $1.6^{+1.0}_{-0.6}$ & \ldots               & $3.6^{+1.6}_{-1.2}$  & $1.9^{+1.1}_{-0.6}$\\
2005am & BL & yes & 0.732(0.003) & 0.043(0.002)        & $0.053(0.017)$ & \ldots              & $3.8^{+1.9}_{-1.1}$ & $1.6^{+0.9}_{-0.5}$ & \ldots               & $3.3^{+1.6}_{-0.8}$  & $2.1^{+0.8}_{-0.6}$\\
2005be & \ldots & no & 0.754(0.018) & 0.029(0.001)     & $<0.039$       & \ldots              & $3.8^{+2.8}_{-1.6}$ & $1.5^{+1.0}_{-0.6}$ & \ldots               & $3.3^{+1.9}_{-1.2}$  & $1.9^{+1.1}_{-0.6}$\\
2005bg & SS & no & 1.002(0.027) & 0.026(0.001)         & $0.078(0.035)$ & \ldots              & $2.5^{+2.4}_{-1.0}$ & $1.5^{+1.0}_{-0.6}$ & \ldots               & $2.3^{+1.3}_{-0.9}$  & $1.9^{+0.9}_{-0.6}$\\
2005bl & CL & no & 0.394(0.013) & 0.025(0.001)         & $0.257(0.050)$ & \ldots              & $1.7^{+1.1}_{-0.8}$ & $1.5^{+1.1}_{-0.5}$ & \ldots               & $1.9^{+1.3}_{-0.7}$  & $1.9^{+0.7}_{-0.6}$\\
2005bo & N & no & 0.846(0.008) & 0.040(0.001)          & $0.327(0.026)$ & $3.5^{+1.8}_{-1.2}$ & $2.3^{+1.0}_{-0.7}$ & $2.0^{+0.8}_{-0.6}$ & $3.3^{+2.0}_{-1.0}$  & $2.6^{+1.1}_{-0.7}$  & $2.3^{+0.7}_{-0.6}$\\
2005el & N & no & 0.834(0.003) & 0.098(0.001)          & $0.015(0.012)$ & \ldots              & $3.9^{+2.1}_{-1.4}$ & $1.5^{+1.0}_{-0.6}$ & \ldots               & $3.5^{+1.6}_{-1.1}$  & $2.0^{+0.9}_{-0.6}$\\
2005eq & SS & no & 1.241(0.008) & 0.063(0.003)         & $0.044(0.024)$ & \ldots              & $2.9^{+2.7}_{-0.9}$ & $1.7^{+1.0}_{-0.6}$ & \ldots               & $3.5^{+1.7}_{-1.1}$  & $2.0^{+1.3}_{-0.5}$\\
2005hc & N & no & 1.191(0.006) & 0.028(0.001)          & $0.049(0.019)$ & \ldots              & $3.8^{+2.4}_{-1.1}$ & $1.8^{+1.7}_{-0.5}$ & \ldots               & $3.8^{+1.7}_{-0.9}$  & $2.2^{+1.4}_{-0.5}$\\
2005hj & \ldots & no & 1.268(0.015) & 0.033(0.001)     & $0.066(0.024)$ & \ldots              & $4.3^{+2.7}_{-1.1}$ & $2.1^{+5.8}_{-0.4}$ & \ldots               & $3.7^{+1.1}_{-0.9}$  & $2.6^{+4.9}_{-0.4}$\\
2005iq & \ldots & no & 0.871(0.004) & 0.019(0.001)     & $0.040(0.015)$ & \ldots              & $4.0^{+2.8}_{-1.3}$ & $1.6^{+1.0}_{-0.5}$ & \ldots               & $3.6^{+1.6}_{-1.0}$  & $2.1^{+0.9}_{-0.6}$\\
2005ir & \ldots & \ldots & 1.120(0.021) & 0.026(0.001) & $0.075(0.025)$ & \ldots              & $2.7^{+2.5}_{-1.1}$ & $1.5^{+1.0}_{-0.6}$ & \ldots               & $2.5^{+1.1}_{-1.1}$  & $1.9^{+0.9}_{-0.6}$\\
2005kc & N & no & 0.898(0.006) & 0.114(0.002)          & $0.310(0.026)$ & $2.1^{+0.5}_{-0.3}$ & $2.1^{+0.4}_{-0.3}$ & $1.8^{+0.4}_{-0.3}$ & $2.6^{+0.4}_{-0.3}$  & $2.5^{+0.4}_{-0.3}$  & $2.3^{+0.3}_{-0.2}$\\
2005ke & CL & no & 0.419(0.003) & 0.020(0.002)         & $0.263(0.033)$ & $0.4^{+0.5}_{-0.2}$ & $0.4^{+0.4}_{-0.2}$ & $0.3^{+0.4}_{-0.2}$ & $0.9^{+0.5}_{-0.2}$  & $0.8^{+0.4}_{-0.2}$  & $0.8^{+0.3}_{-0.2}$\\
2005ki & N & no & 0.824(0.003) & 0.027(0.001)          & $0.016(0.013)$ & \ldots              & $3.6^{+2.9}_{-1.3}$ & $1.5^{+0.9}_{-0.6}$ & \ldots               & $3.4^{+1.6}_{-1.3}$  & $1.9^{+0.9}_{-0.6}$\\
2005ku & N & yes & 1.189(0.044) & 0.046(0.001)         & $0.124(0.048)$ & \ldots              & $2.7^{+1.8}_{-1.3}$ & $1.4^{+1.2}_{-0.4}$ & \ldots               & $2.4^{+1.1}_{-1.0}$  & $1.9^{+0.9}_{-0.7}$\\
2005lu & \ldots & no & 1.128(0.033) & 0.022(0.001)     & $0.247(0.047)$ & \ldots              & $2.2^{+1.2}_{-0.9}$ & $1.5^{+1.1}_{-0.5}$ & \ldots               & $2.5^{+1.1}_{-0.9}$  & $1.9^{+0.9}_{-0.6}$\\
2005mc & \ldots & \ldots & 0.642(0.024) & 0.038(0.001) & $0.212(0.047)$ & \ldots              & $2.2^{+1.1}_{-0.9}$ & $1.6^{+0.9}_{-0.6}$ & \ldots               & $2.5^{+1.1}_{-1.0}$  & $1.9^{+0.9}_{-0.6}$\\
2005na & N & no & 0.958(0.007) & 0.068(0.003)          & $0.061(0.022)$ & \ldots              & $1.9^{+2.0}_{-0.8}$ & $1.3^{+0.8}_{-0.6}$ & \ldots               & $2.5^{+1.8}_{-0.8}$  & $1.6^{+0.8}_{-0.5}$\\
2006D & N & yes & 0.811(0.003) & 0.039(0.001)          & $0.134(0.025)$ & $2.5^{+1.5}_{-0.6}$ & $2.5^{+1.2}_{-0.6}$ & $1.7^{+0.6}_{-0.4}$ & $2.8^{+1.2}_{-0.5}$  & $2.5^{+0.8}_{-0.5}$  & $2.1^{+0.6}_{-0.4}$\\
2006X & BL & yes & 0.970(0.006) & 0.023(0.001)         & $1.360(0.026)$ & $1.3^{+0.1}_{-0.1}$ & $1.3^{+0.1}_{-0.1}$ & $1.2^{+0.1}_{-0.1}$ & $1.8^{+0.1}_{-0.1}$  & $1.8^{+0.1}_{-0.1}$  & $1.7^{+0.1}_{-0.1}$\\
2006ax & N & no & 0.985(0.004) & 0.041(0.002)          & $0.016(0.015)$ & \ldots              & $3.0^{+2.6}_{-1.3}$ & $<2.0$              & \ldots               & $2.9^{+1.7}_{-1.2}$  & $1.7^{+0.9}_{-0.6}$\\
2006bd & CL & no & 0.322(0.021) & 0.023(0.001)         & $0.049(0.055)$ & \ldots              & $1.1^{+1.2}_{-0.5}$ & $1.2^{+0.8}_{-0.6}$ & \ldots               & $1.7^{+1.4}_{-0.7}$  & $1.7^{+0.7}_{-0.6}$\\
2006bh & \ldots & no & 0.800(0.003) & 0.023(0.001)     & $0.037(0.013)$ & \ldots              & $4.3^{+2.5}_{-1.3}$ & $1.7^{+1.0}_{-0.6}$ & \ldots               & $3.8^{+1.8}_{-1.0}$  & $2.1^{+1.1}_{-0.5}$\\
2006br & BL & yes & 0.908(0.029) & 0.020(0.001)        & $0.896(0.050)$ & $1.9^{+0.3}_{-0.2}$ & $1.9^{+0.3}_{-0.2}$ & $1.8^{+0.2}_{-0.2}$ & $2.5^{+0.2}_{-0.2}$  & $2.4^{+0.2}_{-0.2}$  & $2.4^{+0.2}_{-0.2}$\\
2006ef & BL & yes & 0.836(0.020) & 0.020(0.001)        & $0.035(0.032)$ & \ldots              & $3.8^{+2.8}_{-1.6}$ & $1.5^{+1.0}_{-0.6}$ & \ldots               & $3.4^{+1.7}_{-1.2}$  & $1.9^{+1.0}_{-0.6}$\\
2006ej & BL & yes & 0.820(0.019) & 0.030(0.001)        & $0.059(0.030)$ & \ldots              & $2.8^{+2.0}_{-0.9}$ & $1.6^{+0.9}_{-0.6}$ & \ldots               & $3.2^{+1.6}_{-1.0}$  & $2.0^{+0.9}_{-0.6}$\\
2006eq & CL & no & 0.621(0.025) & 0.042(0.001)         & $0.110(0.045)$ & \ldots              & $2.4^{+1.0}_{-0.8}$ & $1.7^{+0.9}_{-0.5}$ & \ldots               & $2.5^{+1.0}_{-0.7}$  & $2.1^{+0.7}_{-0.5}$\\
2006et & N & no & 1.102(0.009) & 0.017(0.001)          & $0.254(0.025)$ & $1.6^{+0.5}_{-0.4}$ & $1.6^{+0.5}_{-0.3}$ & $1.2^{+0.4}_{-0.3}$ & $2.1^{+0.5}_{-0.3}$  & $1.9^{+0.5}_{-0.3}$  & $1.7^{+0.4}_{-0.3}$\\
2006ev & \ldots & no & 0.839(0.019) & 0.076(0.002)     & $0.199(0.045)$ & $3.1^{+2.0}_{-0.6}$ & $2.7^{+0.8}_{-0.5}$ & $2.0^{+0.7}_{-0.4}$ & \ldots               & $3.0^{+0.9}_{-0.5}$  & $2.5^{+0.6}_{-0.4}$\\
2006fw & N & no & 0.895(0.046) & 0.028(0.001)          & $0.262(0.041)$ & \ldots              & $3.7^{+2.8}_{-1.0}$ & $1.8^{+3.8}_{-0.4}$ & \ldots               & $3.0^{+1.3}_{-0.8}$  & $2.2^{+1.4}_{-0.5}$\\
2006gj & CL & no & 0.658(0.007) & 0.070(0.002)         & $0.246(0.025)$ & $3.0^{+0.9}_{-0.5}$ & $2.8^{+0.6}_{-0.4}$ & $2.2^{+0.6}_{-0.4}$ & $3.2^{+0.8}_{-0.4}$  & $3.0^{+0.6}_{-0.4}$  & $2.6^{+0.5}_{-0.3}$\\
2006gt & CL & no & 0.559(0.007) & 0.032(0.001)         & $0.040(0.016)$ & \ldots              & $3.4^{+2.3}_{-1.1}$ & $1.7^{+1.2}_{-0.5}$ & \ldots               & $3.8^{+1.8}_{-1.0}$  & $2.1^{+1.2}_{-0.5}$\\
2006hb & \ldots & no & 0.661(0.004) & 0.024(0.001)     & $0.029(0.018)$ & \ldots              & $2.8^{+1.9}_{-1.1}$ & $1.5^{+0.9}_{-0.6}$ & \ldots               & $3.3^{+1.7}_{-1.1}$  & $1.9^{+0.9}_{-0.6}$\\
2006hx & SS & no & 0.897(0.022) & 0.026(0.001)         & $0.210(0.046)$ & \ldots              & $2.4^{+2.1}_{-0.9}$ & $1.4^{+0.7}_{-0.5}$ & $1.6^{+1.6}_{-0.4}$  & $2.1^{+1.4}_{-0.6}$  & $1.7^{+0.6}_{-0.5}$\\
2006is & N & yes & 1.140(0.032) & 0.029(0.001)         & $<0.024$       & \ldots              & $2.5^{+2.2}_{-1.2}$ & $1.2^{+1.2}_{-0.4}$ & \ldots               & $3.2^{+1.6}_{-1.5}$  & $1.7^{+0.9}_{-0.7}$\\
2006kf & CL & no & 0.733(0.004) & 0.210(0.002)         & $0.032(0.011)$ & \ldots              & $5.0^{+2.3}_{-1.4}$ & $1.7^{+1.6}_{-0.5}$ & \ldots               & $4.1^{+1.7}_{-1.1}$  & $2.2^{+1.3}_{-0.5}$\\
2006lu & \ldots & no & 1.054(0.025) & 0.099(0.002)     & $0.028(0.027)$ & \ldots              & $3.9^{+2.6}_{-1.5}$ & $1.4^{+1.2}_{-0.5}$ & \ldots               & $3.4^{+1.9}_{-1.2}$  & $1.9^{+1.0}_{-0.6}$\\
2006mr & CL & no & 0.260(0.004) & 0.018(0.001)         & $0.089(0.039)$ & \ldots              & $2.8^{+1.1}_{-0.7}$ & $2.1^{+1.4}_{-0.5}$ & \ldots               & $2.9^{+1.1}_{-0.6}$  & $2.5^{+0.8}_{-0.5}$\\
2006ob & \ldots & no & 0.741(0.007) & 0.029(0.001)     & $0.045(0.013)$ & \ldots              & $5.0^{+2.3}_{-1.5}$ & $1.8^{+2.0}_{-0.4}$ & \ldots               & $4.2^{+1.5}_{-1.1}$  & $2.3^{+1.5}_{-0.5}$\\
2006os & CL & yes & 0.937(0.023) & 0.125(0.005)        & $0.471(0.050)$ & $1.6^{+0.4}_{-0.3}$ & $1.6^{+0.4}_{-0.3}$ & $1.4^{+0.4}_{-0.2}$ & $2.1^{+0.4}_{-0.2}$  & $2.0^{+0.3}_{-0.2}$  & $1.9^{+0.3}_{-0.2}$\\
2006py & \ldots & no & 0.949(0.025) & 0.052(0.001)     & $0.154(0.028)$ & \ldots              & $2.1^{+1.1}_{-1.0}$ & $1.5^{+0.9}_{-0.6}$ & \ldots               & $2.4^{+1.1}_{-1.0}$  & $1.8^{+1.1}_{-0.5}$\\
2007A & N & no & 1.003(0.011) & 0.063(0.002)           & $0.259(0.024)$ & $2.3^{+0.7}_{-0.4}$ & $2.2^{+0.5}_{-0.4}$ & $1.8^{+0.5}_{-0.3}$ & $2.7^{+0.6}_{-0.3}$  & $2.5^{+0.5}_{-0.3}$  & $2.2^{+0.4}_{-0.3}$\\
2007N & CL & no & 0.297(0.007) & 0.034(0.002)          & $0.350(0.052)$ & $1.4^{+0.6}_{-0.4}$ & $1.1^{+0.4}_{-0.3}$ & $1.1^{+0.4}_{-0.3}$ & $1.8^{+0.6}_{-0.3}$  & $1.5^{+0.4}_{-0.2}$  & $1.5^{+0.4}_{-0.2}$\\
2007S & SS & no & 1.121(0.010) & 0.022(0.002)          & $0.478(0.026)$ & $1.6^{+0.3}_{-0.2}$ & $1.6^{+0.2}_{-0.2}$ & $1.4^{+0.2}_{-0.2}$ & $2.1^{+0.2}_{-0.2}$  & $1.9^{+0.2}_{-0.2}$  & $1.9^{+0.2}_{-0.2}$\\
2007af & BL & no & 0.926(0.003) & 0.034(0.001)         & $0.178(0.024)$ & $1.7^{+0.7}_{-0.4}$ & $1.8^{+0.7}_{-0.4}$ & $2.1^{+0.6}_{-0.5}$ & $2.1^{+0.7}_{-0.4}$  & $2.1^{+0.6}_{-0.4}$  & $1.8^{+0.5}_{-0.3}$\\
2007ai & SS & no & 1.251(0.011) & 0.286(0.004)         & $0.160(0.025)$ & \ldots              & $4.1^{+1.0}_{-0.6}$ & $3.4^{+1.1}_{-0.7}$ & \ldots               & $4.3^{+0.9}_{-0.5}$  & $3.8^{+2.6}_{-0.5}$\\
2007as & BL & yes & 0.881(0.004) & 0.123(0.001)        & $0.050(0.011)$ & \ldots              & $4.7^{+2.7}_{-1.1}$ & $2.1^{+6.2}_{-0.3}$ & \ldots               & $4.0^{+1.2}_{-0.8}$  & $3.0^{+7.8}_{-0.3}$\\
2007ax & CL & no & 0.360(0.010) & 0.045(0.001)         & $0.213(0.049)$ & $3.0^{+1.7}_{-0.7}$ & $2.1^{+0.8}_{-0.5}$ & $1.9^{+0.6}_{-0.4}$ & $3.3^{+1.7}_{-0.6}$  & $2.6^{+0.8}_{-0.4}$  & $2.4^{+0.6}_{-0.4}$\\
2007ba & CL & no & 0.547(0.006) & 0.032(0.002)         & $0.150(0.026)$ & $<0.7$              & $<1.3$              & $0.3^{+0.7}_{-0.1}$ & $0.6^{+0.6}_{-0.2}$  & $0.7^{+0.7}_{-0.2}$  & $0.7^{+0.6}_{-0.2}$\\
2007bc & CL & no & 0.886(0.005) & 0.019(0.001)         & $0.207(0.025)$ & $1.2^{+0.7}_{-0.4}$ & $1.4^{+0.6}_{-0.4}$ & $1.0^{+0.5}_{-0.3}$ & $1.8^{+0.6}_{-0.4}$  & $1.8^{+0.5}_{-0.4}$  & $1.6^{+0.4}_{-0.4}$\\
2007bd & BL & yes & 0.883(0.004) & 0.029(0.001)        & $0.058(0.022)$ & \ldots              & $2.2^{+2.4}_{-0.9}$ & $1.1^{+0.8}_{-0.5}$ & \ldots               & $2.1^{+2.1}_{-0.7}$  & $1.4^{+0.8}_{-0.5}$\\
2007bm & N & no & 0.905(0.008) & 0.035(0.001)          & $0.606(0.025)$ & $1.3^{+0.2}_{-0.2}$ & $1.3^{+0.2}_{-0.2}$ & $1.2^{+0.2}_{-0.1}$ & $1.8^{+0.2}_{-0.1}$  & $1.8^{+0.2}_{-0.1}$  & $1.7^{+0.2}_{-0.1}$\\
2007ca & N & no & 1.060(0.007) & 0.057(0.002)          & $0.350(0.024)$ & $2.9^{+0.4}_{-0.3}$ & $2.6^{+0.4}_{-0.3}$ & $2.3^{+0.4}_{-0.3}$ & $3.1^{+0.4}_{-0.3}$  & $2.9^{+0.4}_{-0.3}$  & $2.7^{+0.3}_{-0.3}$\\
2007hx & \ldots & \ldots & 1.016(0.031) & 0.030(0.001) & $0.266(0.046)$ & $2.5^{+1.9}_{-0.7}$ & $2.3^{+0.9}_{-0.6}$ & $1.8^{+0.7}_{-0.5}$ & $2.7^{+1.4}_{-0.5}$  & $2.5^{+0.8}_{-0.5}$  & $2.2^{+0.6}_{-0.4}$\\
2007if & \ldots & \ldots & 1.241(0.020) & 0.071(0.006) & $<0.051$       & \ldots              & $4.5^{+2.2}_{-1.7}$ & $1.7^{+0.9}_{-0.6}$ & \ldots               & $3.7^{+1.8}_{-1.2}$  & $2.0^{+0.9}_{-0.6}$\\
2007jg & BL & yes & 0.915(0.006) & 0.090(0.002)        & $0.108(0.025)$ & $2.4^{+2.1}_{-0.8}$ & $2.4^{+1.6}_{-0.7}$ & $1.5^{+0.7}_{-0.5}$ & $3.0^{+2.4}_{-0.7}$  & $2.5^{+1.0}_{-0.6}$  & $2.0^{+0.7}_{-0.5}$\\
2007jh & \ldots & no & 0.586(0.014) & 0.090(0.003)     & $0.169(0.034)$ & \ldots              & $1.8^{+1.1}_{-0.9}$ & $1.3^{+0.8}_{-0.6}$ & \ldots               & $2.1^{+1.0}_{-1.0}$  & $1.8^{+0.8}_{-0.7}$\\
2007le & BL & yes & 1.023(0.004) & 0.029(0.001)        & $0.388(0.023)$ & $1.4^{+0.3}_{-0.2}$ & $1.3^{+0.3}_{-0.2}$ & $1.5^{+0.3}_{-0.2}$ & $1.8^{+0.3}_{-0.2}$  & $1.7^{+0.2}_{-0.2}$  & $1.6^{+0.2}_{-0.2}$\\
2007mm & \ldots & \ldots & 0.500(0.020) & 0.031(0.001) & $0.162(0.049)$ & \ldots              & $1.8^{+1.1}_{-0.9}$ & $1.3^{+0.8}_{-0.6}$ & \ldots               & $2.0^{+1.0}_{-1.1}$  & $1.7^{+0.8}_{-0.7}$\\
2007nq & BL & yes & 0.749(0.005) & 0.031(0.001)        & $0.046(0.013)$ & \ldots              & $5.0^{+2.8}_{-1.2}$ & $1.9^{+4.5}_{-0.4}$ & \ldots               & $4.5^{+1.8}_{-1.1}$  & $2.4^{+2.9}_{-0.5}$\\
2007on & CL & no & 0.574(0.003) & 0.010(0.001)         & $<0.007$       & \ldots              & $4.1^{+2.0}_{-1.7}$ & $1.9^{+0.8}_{-0.6}$ & \ldots               & $3.5^{+1.7}_{-1.4}$  & $1.9^{+1.0}_{-0.6}$\\
2008C & SS & no & 0.947(0.024) & 0.072(0.002)          & $0.270(0.046)$ & $1.0^{+0.8}_{-0.3}$ & $1.2^{+0.7}_{-0.4}$ & $2.4^{+0.6}_{-0.7}$ & $1.5^{+0.7}_{-0.3}$  & $1.5^{+0.6}_{-0.3}$  & $1.4^{+0.5}_{-0.3}$\\
2008R & CL & no & 0.597(0.006) & 0.062(0.001)          & $0.009(0.013)$ & \ldots              & $3.8^{+2.5}_{-1.5}$ & $1.5^{+1.0}_{-0.6}$ & \ldots               & $3.2^{+1.8}_{-1.1}$  & $1.8^{+0.9}_{-0.6}$\\
2008bc & N & no & 1.035(0.003) & 0.225(0.004)          & $<0.019$       & \ldots              & $3.1^{+2.4}_{-1.8}$ & $1.2^{+1.0}_{-0.5}$ & \ldots               & $3.1^{+1.4}_{-1.8}$  & $1.6^{+0.9}_{-0.8}$\\
2008bq & N & no & 1.151(0.008) & 0.077(0.002)          & $0.136(0.027)$ & $1.8^{+1.7}_{-0.7}$ & $1.7^{+1.2}_{-0.7}$ & $1.3^{+0.7}_{-0.5}$ & $2.3^{+1.8}_{-0.7}$  & $2.0^{+0.8}_{-0.6}$  & $1.7^{+0.6}_{-0.5}$\\
2008fp & N & no & 1.049(0.005) & 0.169(0.002)          & $0.578(0.024)$ & $1.4^{+0.2}_{-0.2}$ & $1.3^{+0.2}_{-0.2}$ & $1.2^{+0.2}_{-0.1}$ & $1.8^{+0.2}_{-0.1}$  & $1.7^{+0.2}_{-0.1}$  & $1.7^{+0.2}_{-0.1}$\\
2008gp & \ldots & no & 0.974(0.005) & 0.104(0.005)     & $0.098(0.022)$ & $<0.5$              & $<0.7$              & $<1.1$              & $0.5^{+0.6}_{-0.2}$  & $0.7^{+0.8}_{-0.2}$  & $0.7^{+0.7}_{-0.2}$\\
2008hv & N & no & 0.846(0.003) & 0.028(0.001)          & $0.074(0.023)$ & $1.0^{+2.3}_{-0.3}$ & $1.6^{+1.5}_{-0.7}$ & $2.2^{+0.6}_{-0.7}$ & $1.5^{+2.3}_{-0.4}$  & $2.1^{+1.8}_{-0.7}$  & $1.5^{+0.8}_{-0.5}$\\
2008ia & BL & no & 0.837(0.004) & 0.195(0.005)         & $0.066(0.016)$ & \ldots              & $4.0^{+2.0}_{-1.0}$ & $2.4^{+0.6}_{-0.6}$ & \ldots               & $3.8^{+1.6}_{-0.7}$  & $2.4^{+1.2}_{-0.5}$\\
2009F & CL & no & 0.335(0.008) & 0.089(0.002)          & $0.108(0.047)$ & \ldots              & $0.8^{+1.3}_{-0.3}$ & $0.8^{+0.8}_{-0.4}$ & \ldots               & $1.0^{+1.2}_{-0.2}$  & $1.0^{+0.8}_{-0.3}$
\enddata
\tablecomments{Column 1: IAU Name; Column 2: \citet{Branch:2009} spectroscopic classification from \citet{Folatelli:2013}; 
Column 3: whether the SN is high-velocity (HV) \citep{Wang:2009}; Column 4: stretch parameter; Column 5: Milky-Way
foreground $B-V$ color excess from \citet{Schlafly:2011}; Column 6: the host-galaxy $B-V$ color excess; Columns
7-9: the best-fit $R_V$ when using CCM+O for three priors (Uniform, Binned, and Gaussian mixture model, respectively);
Columns 10-12: same as columns 7-9, but using F99.}
\end{deluxetable*}

\clearpage
\begin{deluxetable*}{cllll}
\tablecolumns{5}
\tablecaption{Intrinsic Color Coefficients\label{tab:color_coef}}
\tablehead{
\colhead{Color} & \colhead{$a$} & \colhead{$b$} & \colhead{$c$} & \colhead{$\sigma_{xy}$}  }
\startdata
\sidehead{Cauchy Prior}
$u - B$ &    $-0.66(0.04)$ &    $-0.91(0.29)$   & \phs $4.17(0.7)$  & $0.15$\\
$B - V$ & \phs$0.91(0.03)$ & \phs$0.62(0.20)$   &     $-1.43(0.4)$  & $0.07$\\
$g - r$ &   $ -0.25(0.01)$ &    $-0.12(0.07)$   & \phs $0.99(0.2)$  & $0.05$\\
$r - i$ &    $-0.63(0.01)$ &    $-0.22(0.08)$   & \phs $0.53(0.2)$  & $0.08$\\
$V - Y$ &    $-0.79(0.02)$ &    $-0.66(0.15)$   & \phs $1.01(0.3)$  & $0.06$\\
$V - J$ &    $-0.68(0.03)$ &    $-0.39(0.18)$   & \phs $1.45(0.4)$  & $0.09$\\
$V - H$ &    $-0.91(0.03)$ &    $-0.62(0.20)$   & \phs $1.43(0.4)$  & $0.06$\\
$Y - J$ & \phs$0.11(0.02)$ & \phs$0.27(0.10)$   & \phs $0.44(0.2)$  & $0.10$\\
$J - H$ &    $-0.23(0.01)$ &    $-0.23(0.10)$   &     $-0.03(0.2)$  & $0.09$\\
\sidehead{Cauchy Prior, $s_{BV} > 0.5$}
$u - B$ &     $-0.629(0.040)$ &     $-1.31(0.35)$ & \phs $2.4(1.0)$ & $0.14$\\
$B - V$ & \phs $0.910(0.025)$ & \phs$ 0.87(0.24)$ &     $-0.9(0.6)$ & $0.06$\\
$g - r$ &     $-0.230(0.013)$ &     $-0.22(0.09)$ & \phs $0.3(0.3)$ & $0.05$\\
$r - i$ &     $-0.634(0.014)$ &     $-0.27(0.09)$ & \phs $0.5(0.3)$ & $0.08$\\
$V - Y$ &     $-0.781(0.019)$ &     $-0.88(0.18)$ & \phs $0.2(0.5)$ & $0.06$\\
$V - J$ &     $-0.685(0.023)$ &     $-0.60(0.21)$ & \phs $1.0(0.6)$ & $0.06$\\
$V - H$ &     $-0.910(0.025)$ &     $-0.87(0.24)$ & \phs $0.9(0.6)$ & $0.06$\\
$Y - J$ & \phs $0.096(0.016)$ & \phs$ 0.28(0.09)$ & \phs $0.8(0.3)$ & $0.09$\\
$J - H$ &     $-0.225(0.016)$ &     $-0.27(0.09)$ &     $-0.1(0.3)$ & $0.07$\\
\sidehead{LRS Prior}
$u - B$ &     $-0.54(0.03)$ &     $-0.79(0.17)$ & \phs$ 5.2(0.3)$ & $0.17$\\
$B - V$ & \phs$ 0.82(0.02)$ & \phs$ 0.51(0.11)$ &     $-1.9(0.2)$ & $0.09$\\
$g - r$ &     $-0.22(0.01)$ &     $-0.10(0.06)$ & \phs$ 1.3(0.1)$ & $0.07$\\
$r - i$ &     $-0.61(0.01)$ &     $-0.20(0.08)$ & \phs$ 0.7(0.2)$ & $0.10$\\
$V - Y$ &     $-0.73(0.01)$ &     $-0.60(0.09)$ & \phs$ 1.5(0.2)$ & $0.08$\\
$V - J$ &     $-0.61(0.02)$ &     $-0.29(0.12)$ & \phs$ 2.0(0.2)$ & $0.12$\\
$V - H$ &     $-0.82(0.02)$ &     $-0.51(0.11)$ & \phs$ 1.9(0.2)$ & $0.10$\\
$Y - J$ & \phs$ 0.12(0.02)$ & \phs$ 0.31(0.12)$ & \phs$ 0.6(0.2)$ & $0.13$\\
$J - H$ &     $-0.21(0.02)$ &     $-0.22(0.14)$ &     $-0.1(0.3)$ & $0.14$\\
\sidehead{LRS Prior, $s_{BV} > 0.5$}
$u - B$ &     $-0.50(0.03)$ &     $-0.89(0.16)$ & \phs $4.1(0.6)$ &$ 0.16$\\
$B - V$ & \phs$ 0.81(0.02)$ & \phs$ 0.55(0.11)$ &     $-1.7(0.4)$ &$ 0.09$\\
$g - r$ &     $-0.20(0.01)$ &     $-0.14(0.06)$ & \phs $0.9(0.2)$ &$ 0.06$\\
$r - i$ &     $-0.61(0.02)$ &     $-0.20(0.08)$ & \phs $0.7(0.3)$ &$ 0.10$\\
$V - Y$ &     $-0.71(0.02)$ &     $-0.66(0.08)$ & \phs $0.9(0.3)$ &$ 0.08$\\
$V - J$ &     $-0.60(0.02)$ &     $-0.32(0.09)$ & \phs $1.8(0.3)$ &$ 0.08$\\
$V - H$ &     $-0.81(0.02)$ &     $-0.55(0.11)$ & \phs $1.7(0.4)$ &$ 0.09$\\
$Y - J$ & \phs$ 0.11(0.02)$ & \phs$ 0.34(0.10)$ & \phs $0.9(0.4)$ &$ 0.09$\\
$J - H$ &     $-0.21(0.02)$ &     $-0.23(0.11)$ &     $-0.1(0.4)$ &$ 0.11$
\enddata
\tablecomments{Column 1: The pseudo color at maximum; Columns 2-4: the coefficients of the 
polynomial $a + b\left(s_{BV}-1\right) + c\left(s_{BV}-1\right)^2$; Column 5: the intrinsic
scatter for each color.}
\end{deluxetable*}
\clearpage

\begin{deluxetable*}{lccc}
\tablewidth{0pc}
\tablecolumns{4}
\tablecaption{Global $R_V$ Parameter for Different Reddening Laws and Samples\label{tab:Rv_global}}
\tablehead{
\colhead{Dataset} & \colhead{$R_V$ (CCM+O)} & \colhead{$R_V$ (F99)} & \colhead{$p$ (Goobar)}  }
\startdata
all objects         & 1.27(0.07) & 1.81(0.06) & $-2.20(0.07)$\\
$s_{BV}>0.5$        & 1.32(0.07) & 1.85(0.07) & $-2.14(0.07)$\\
$E(B-V)<0.5$        & 1.76(0.18) & 2.15(0.16) & $-1.63(0.14)$\\
$u$-band excluded, all objects              & 1.31(0.07) & 1.86(0.06) & $-2.26(0.08)$\\
$u$-band excluded, $s_{BV}>0.5$     & 1.36(0.07) & 1.90(0.06) & $-2.19(0.08)$\\
$u$-band excluded ,$E(B-V)<0.5$     & 1.78(0.19) & 2.18(0.16) & $-1.69(0.15)$
\enddata
\tablecomments{Column 1: description of sub-samples; Column 2: value of $R_V$ when assuming
a CCM+O reddening law; Column 3: value of $R_V$ when assuming an F99 reddening law; and
Column 4: power law index when assuming a \citet{Goobar:2008} reddening law.}
\end{deluxetable*}

\begin{deluxetable*}{lllcccccc}
\tablewidth{0pc}
\tablecolumns{4}
\tablecaption{Inferred Milky-Way Color Excesses\label{tab:MW_EBV}}
\tablehead{
\colhead{SN} & \colhead{$E(B-V)_{S11}$} & \colhead{$E(B-V)$} & \colhead{$R_{V}$}}
\startdata
2006kf & 0.210(0.002) & $0.19(0.02)$ & $4.1^{+1.1}_{-0.5}$ \\
2008bc & 0.225(0.004) & $0.25(0.03)$ & $2.1^{+0.5}_{-0.3}$ \\
2008ia & 0.195(0.005) & $0.24(0.03)$ & $3.9^{+0.6}_{-0.4}$
\enddata
\tablecomments{Column 1: IAU Name; Column 2: Milky-Way $E(B-V)$ color excesses
from \citet{Schlafly:2011}; Column 3: Milky-Way $E(B-V)$ color excesses 
inferred from SN colors; Column 4: Milky-Way $R_V$ inferred from
SN colors.}
\end{deluxetable*}

\begin{deluxetable*}{lllllll}
\tablewidth{0pc}
\tablecolumns{7}
\tablecaption{Gaussian Mixture Model Hyper-Parameters\label{tab:Ngauss}}
\tablehead{
\colhead{Dataset} & \colhead{$\pi_1$} & \colhead{$\mu_1$} & \colhead{$\sigma_1$} & \colhead{$\pi_2$} & \colhead{$\mu_2$} & \colhead{$\sigma_2$}  }
\startdata
\cutinhead{CCM+O Reddening Law}
All objects                          & 0.95 & 1.6 (0.2) & 0.71 (0.04)& 0.05 & 5.7 (0.5)& 1.4 (0.1)\\
$s_{BV} > 0.5$                       & 0.97 & 2.2 (0.2) & 0.76 (0.04)& 0.03 & 3.3 (0.5)& 1.4 (0.2)\\
$B-V < 0.5$                          & 0.95 & 2.5 (0.3) & 0.88 (0.05)& 0.05 & 5.6 (0.6)& 1.4 (0.2)\\
$u$ band excluded, all objects       & 0.94 & 1.6 (0.2) & 0.72 (0.04)& 0.06 & 6.2 (0.4)& 1.5 (0.2)\\
$u$ band excluded, $s_{BV} > 0.5$    & 0.95 & 2.2 (0.3) & 0.78 (0.05)& 0.05 & 4.1 (0.4)& 1.4 (0.1)\\
$u$ band excluded, $B-V < 0.5$       & 0.95 & 2.5 (0.3) & 0.92 (0.06)& 0.05 & 5.4 (0.6)& 1.4 (0.1)\\
\cutinhead{F99 Reddening Law}
All objects                          & 0.96 & 2.0(0.2) & 0.72(0.03) & 0.04 & 6.0(0.5) & 1.5(0.4)\\
$s_{BV} > 0.5$                       & 0.97 & 2.1(0.2) & 0.75(0.04) & 0.03 & 5.7(0.6) & 1.5(0.4)\\
$B-V < 0.5$                          & 0.96 & 2.3(0.3) & 0.84(0.05) & 0.04 & 6.3(0.6) & 1.5(0.6)\\
$u$ band excluded, all objects       & 0.96 & 2.0(0.2) & 0.78(0.03) & 0.04 & 6.0(0.6) & 1.5(0.3)\\
$u$ band excluded, $s_{BV} > 0.5$    & 0.95 & 2.1(0.2) & 0.81(0.03) & 0.05 & 6.1(0.5) & 1.5(0.1)\\
$u$ band excluded, $B-V < 0.5$       & 0.96 & 2.3(0.3) & 0.93(0.05) & 0.04 & 5.8(0.7) & 1.5(0.4)
\enddata
\tablecomments{Column 1: description of sub-samples; Columns 2-4: fraction, mean, and standard
deviation of the first Gaussian component; Columns 5-7: fraction, mean, and standard
deviation of the second Gaussian component.}
\end{deluxetable*}
\clearpage
\begin{deluxetable*}{lllllc}
\tablewidth{0pc}
\tablecolumns{6}
\tablecaption{Binned Prior Hyper-Parameters\label{tab:Rv_binned}}
\tablehead{ \colhead{}             & \multicolumn{2}{c}{CCM+O} & \multicolumn{2}{c}{F99} & \colhead{} \\
\cline{2-3} \cline{4-5}\\
\colhead{$E(B-V)$ bin} & \colhead{$\mu$} & \colhead{$\sigma$} & \colhead{$\mu$} & \colhead{$\sigma$} & \colhead{\# of SNe}  }
\startdata
\sidehead{All SNe}
$E(B-V)_{host} < 0.1$       & 4.3(1.0) & 1.8(0.8) &3.6(0.8) & 1.3(0.4) & 40\\
$0.1 < E(B-V)_{host} < 0.3$ & 3.2(0.8) & 1.8(0.7) &2.5(0.3) & 1.1(0.2) & 30\\
$0.3 < E(B-V)_{host} < 0.4$ & 2.2(0.4) & 1.1(0.3) &2.2(0.6) & 1.2(0.5) & 4\\
$0.4 < E(B-V)_{host} < 0.7$ & 1.7(0.4) & 1.1(0.3) &1.8(0.5) & 1.0(0.5) & 5\\
$E(B-V)_{host} > 0.7$       & 1.5(0.5) & 1.0(0.5) &2.0(0.9) & 1.3(0.9) & 3\\
\sidehead{$s_{BV} > 0.5$}
$E(B-V)_{host} < 0.1$       & 4.6(1.0) & 1.5(0.6) &4.1(0.9) & 1.5(0.6) & 36\\
$0.1 < E(B-V)_{host} < 0.3$ & 3.2(0.9) & 1.4(0.7) &3.3(0.7) & 1.2(0.5) & 26\\
$0.3 < E(B-V)_{host} < 0.4$ & 2.4(0.5) & 1.2(0.4) &2.6(0.4) & 1.1(0.3) & 2\\
$0.4 < E(B-V)_{host} < 0.7$ & 2.2(0.5) & 1.3(0.4) &2.4(0.5) & 1.3(0.4) & 5\\
$E(B-V)_{host} > 0.7$       & 1.4(0.5) & 1.0(0.4) &1.8(0.5) & 1.0(0.4) & 3
\enddata
\tablecomments{Column 1: The $E(B-V)$ color excess bin; Column 2: the mean value of $R_V$ for the bin
when using CCM+O; Column 3: the standard deviation for the bin when using CCM+O; Column 4:
the mean value of $R_V$ for the bin when using F99; Column 5: the standard deviation for the 
bin when using F99; Column 6: number of objects in each bin.}
\end{deluxetable*}
\end{document}